# Smart Healthcare in the Age of AI: Recent Advances, Challenges, and Future Prospects

**MAHMOUD NASR[1], MD. MILON ISLAM[1], SHADY SHEHATA[1], FAKHRI KARRAY[1] (FELLOW, IEEE), AND YURI QUINTANA[2]**

[1]Centre for Pattern Analysis and Machine Intelligence, Department of Electrical and Computer Engineering, University of Waterloo, ON, Canada N2L 3G1
[2]Division of Clinical Informatics, Beth Israel Deaconess Medical Center and Harvard Medical School, Boston, MA, USA

Corresponding author: Mahmoud Nasr (mmmnasrm@uwaterloo.ca) and Md. Milon Islam (e-mail: milonislam@uwaterloo.ca).

**ABSTRACT** The significant increase in the number of individuals with chronic ailments (including the elderly and disabled) has dictated an urgent need for an innovative model for healthcare systems. The evolved model will be more personalized and less reliant on traditional brick-and-mortar healthcare institutions such as hospitals, nursing homes, and long-term healthcare centers. The smart healthcare system is a topic of recently growing interest and has become increasingly required due to major developments in modern technologies, especially in artificial intelligence (AI) and machine learning (ML). This paper is aimed to discuss the current state-of-the-art smart healthcare systems highlighting major areas like wearable and smartphone devices for health monitoring, machine learning for disease diagnosis, and the assistive frameworks, including social robots developed for the ambient assisted living environment. Additionally, the paper demonstrates software integration architectures that are very significant to create smart healthcare systems, integrating seamlessly the benefit of data analytics and other tools of AI. The explained developed systems focus on several facets: the contribution of each developed framework, the detailed working procedure, the performance as outcomes, and the comparative merits and limitations. The current research challenges with potential future directions are addressed to highlight the drawbacks of existing systems and the possible methods to introduce novel frameworks, respectively. This review aims at providing comprehensive insights into the recent developments of smart healthcare systems to equip experts to contribute to the field.

## I. INTRODUCTION

With projections of 22% of the population reaching the age 60 or more by 2050 [1], people affected by chronic diseases are growing along with health-related emergencies, therefore resulting in a higher pressure on the healthcare industry [2], [3]. With a decline in the ratio between working-age people, there are fewer professional healthcare workers to care for the increase in demand. Besides, the cost of said health care, medications, and medical devices continuously soar, making it harder to cover such costs for the average citizen as the need for more caregivers and healthcare facilities increases to with-stand the increase in demand [4]. Combined, these conditions call for cheaper, more inclusive, and better health care solutions. A great candidate for such a situation is utilizing the recent advancements in smart and miniaturized sensors, communication technologies, and artificial intelligence to provide technological solutions at an affordable price to the broadest range of the population without sacrificing the quality of care.

The Internet of Things (IoT) [5]–[7] has been steadily increasing in popularity over the past years. Due to the advancements in communication technologies and speed of data transfer, the ability to transmit large amounts of data has grown drastically. In addition, more robust and advanced storage and processing capabilities provided by big data analytics [8], [9], and cloud technologies [10], [11] opened the doors for new applications and markets for IoT in real-time analytics and predictive modelling. As a result of the earlier described advancements with combining g smart sensors, actuators, and data analytics in an IoT environment for real-time and continuous healthcare brings great promise to the healthcare industry. The field, called the Internet of Health Things (IoHT) [12], [13] or the Internet of Medical Things (IoMT) [14], [15] offers the potential of transforming the healthcare paradigm. The method which is pursued in research and practice can be divided into several areas. First, smart sensors are combined in the IoMT environment to continuously monitor health vitals using wearable devices [16]–[21] and smartphone solutions such as those proposed by [22]–[25]. After the data is collected using smart sensors, machine learning techniques interpret the data and present predictive analytics such as predicting illnesses [26]–[28]. In addition, other algorithms are used to keep track of chronic conditions such as diabetes [29]–[31] and heart diseases [32], [33] and detect abnormalities in the patient's health.

One of the main drivers for smart healthcare adoption is the increasing ratio of older adults in societies worldwide. Consequently, ambient assisted living focuses on creating environments for older adults that integrate smart healthcare techniques for better care without human intervention. Given that around 90% of older adults prefer staying at their own homes, many solutions are based on smart home systems such as those proposed in [34]–[37]. User studies [36], [38], [39], however, uncover the importance of including robotic agents capable of social interactions with the user to provide both psychological and physical assistance to older adults. Several studies tackle using robotic agents for the care of older adults [39]–[41], while others propose taking a step further by integrating such robots in the ambient assisted living (AAL) environments with other smart sensors [42]–[44]. Integrating various sensors, actuators, and user interfaces requires rigorous work on a scalable and personalization to different user needs. Therefore, studies such as [42], [45], [46] attempt to formalize different architecture to tackle this problem and create an integrated smart healthcare f

The goal of this paper is to explore the state-of-the-art smart healthcare systems that highlight the significant areas of research, including wearable and smartphone-based health monitoring, machine learning for predictive analytics, and assistive frameworks developed for assisted living environments, including social robots. The main contributions of the paper can be summarized as follows:

- Provide a systemic review of state-of-the-art research in smart medical devices, machine learning for disease prediction, AAL, and software architectures.
- Compare approaches to each problem, highlight their advantages and challenges, and present recommendations for improvement in future studies.
- Present a holistic overview of the smart healthcare field to provide a complete view of how technologies in different areas can be combined to accelerate smart healthcare.

The overall workflow of the reviewed systems is shown in Figure 1. In the Figure, the comprehensive review of smart healthcare systems is divided into three major areas, namely: health monitoring, disease diagnosis, and supportive devices in AAL. Additionally, the software integration architectures are described in this review. The health monitoring prototypes are divided according to wearable devices or smartphones, mentioned in Section II. Three major diseases, like COVID-19, heart disease, and diabetes detection frameworks based on machine learning algorithms, are demonstrated throughout Section III. Section IV discusses the assistive prototypes in AAL, the supportive tools in smart homes, and social robots. The major directions in creating software architecture for smart healthcare are discussed in section V. Section VI presents an open discussion of the reviewed studies and guidelines for areas of future work. Finally, section VII concludes the study and summarizes the main key points.

## II. HEALTHCARE MONITORING DEVICES IN IOT

It has been commonly recognized as the Internet of Things is used as a possible solution to relieve the stresses on healthcare infrastructures and has become a prominent research issue in recent times [47]–[49]. Health monitoring can be ensured through wearable sensors and smartphone applications. A broad pipeline of a health monitoring system based on wearable devices is shown in Figure 2. Various sensors collect data from patients and transfer it to a cloud server for processing through Wi-Fi or LoRa gateway. The health status of the patients is monitored through physicians and their family members through several user interface tools like web or mobile applications. In some cases, emergency services are deployed to handle critical situations of the patients. The significant developments of health monitoring through IoT are described in this section, digging deeper into the details of implementation and technologies utilized.

### A. WEARABLE DEVICES FOR HEALTH MONITORING

Smart healthcare services using wearable sensors provide an appropriate and cheaper alternative to the costly hospital environment [16]–[19]. These systems enable medical professionals to screen the significant symptoms of the patients, evaluate the general health of the users, and detect abnormalities remotely.

Recently, Islam et al. [20] developed a smart healthcare system that monitored patients' health using five sensors: two sensors (heart rate sensor and body temperature sensor (LM35)) for patient condition monitoring and three sensors (room temperature sensor (DHT11), CO sensor (MQ-9), and $CO_2$ sensor (MQ-135)) for detection of the living environment condition. Here, the processing device is ESP32, and Wi-Fi is used as communication media to transfer data from the patient's side to a web server. However, the developed prototype is not adequately manufactured for application purposes. Afterward, the authors of [50] proposed a wearable remote healthcare monitoring (RHM) framework that can monitor the heart rate (HR), body temperature and detect falls. The proposed scheme used NodeMCU as a processing device and heartbeat sensor, body temperature sensor (LM35), and accelerometer (MPU 6050) as sensing elements. The ThingSpeak web service is used for data visualization to aid physicians in monitoring the patients from remote locations with a hand-held device as a prototype. In another study, the authors of [51] demonstrated a healthcare monitoring framework utilizing the concept of IoT and cloud computing. The prototype used an HR sensor, ECG, SpO2, and body temperature sensors for monitoring the corresponding heart rate, ECG, oxygen saturation, and body temperature of the patients, respectively. Two microcontrollers (Arduino and NodeMCU) were utilized to collect data from sensors and transfer data from edge devices to the cloud using Wi-Fi. Blynk cloud service was used to monitor the patients' health parameters from remote locations. However, the system is not able to handle the emergencies of the patients.

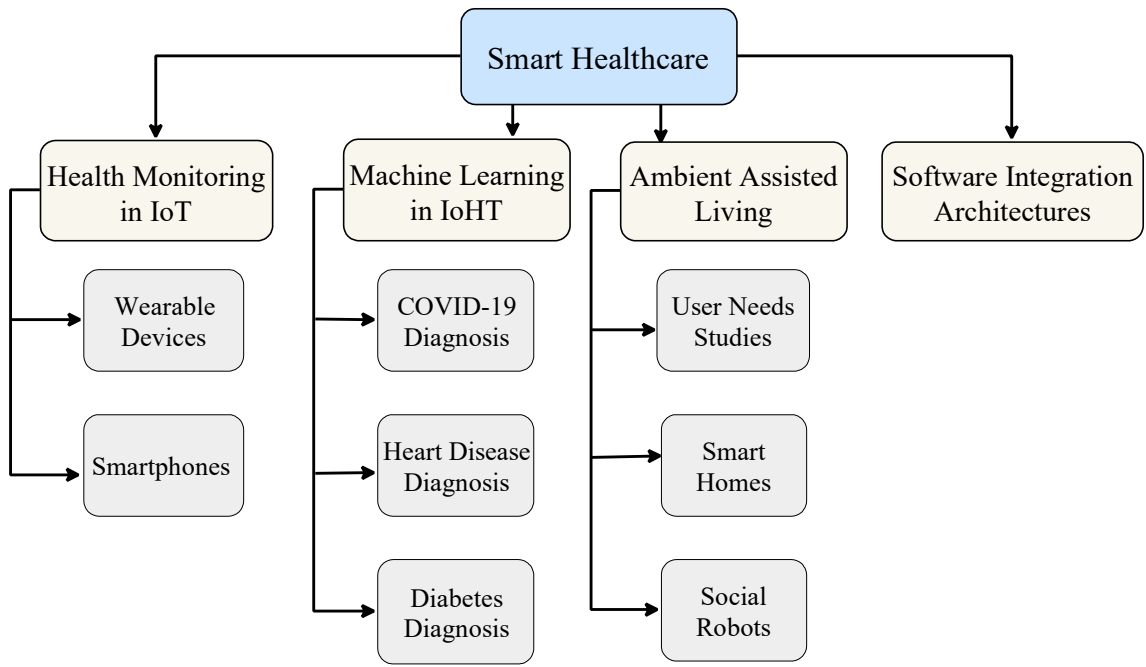

**FIGURE 1.** Overall workflow of the reviewed systems of smart healthcare.

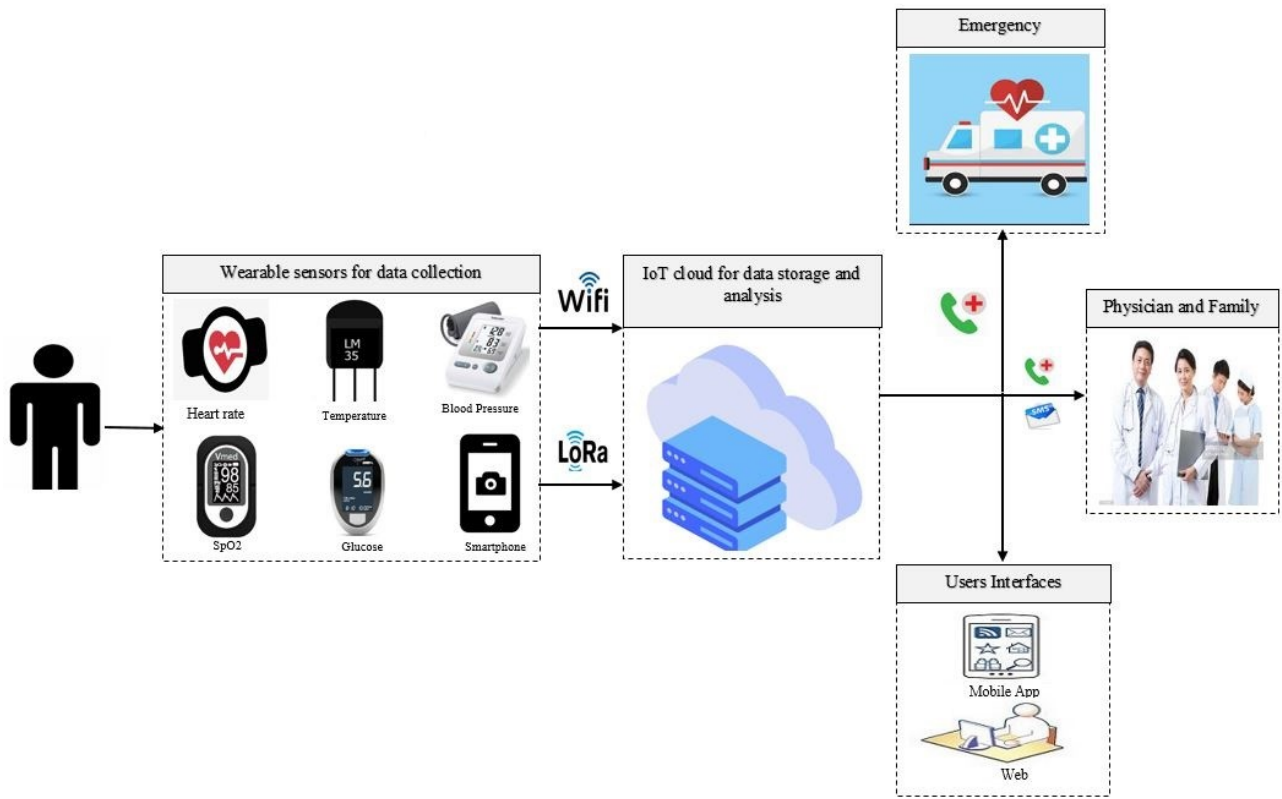

**FIGURE 2.** A general pipeline of a health monitoring system based on wearable devices.

Chigozirim et al. [52] introduced a patient monitoring prototype that allows the doctors to monitor the patient's status through the developed tool using IoT. In this system, HR and body temperature are monitored through pulse and body temperature sensors. ATmega328P microcontroller and NodeMCU are used as edge devices, and the collected data from sensors are transmitted to the Internet through a Wi-Fi connection. In addition, the doctors can check the patient's condition using LCD. However, the developed system has not been entirely fabricated for real-time tests. In another research, Mohapatra et al. [53] demonstrated a smart healthcare management framework in IoT and cloud services to ensure the patient's condition monitoring from remote locations. The proposed scheme utilized HR and temperature sensors to perceive the data from patients, and an Arduino is used as an edge device connected to sensors and used to transmit data to the cloud through Wi-Fi. The doctors can easily connect with the cloud server through any internet-connected devices to check the patient status and suggest proper medications in an emergency. However, no security concerns are mentioned here for the transmitted data. Further, Swaroop et al. [54] developed a framework for basic symptoms monitoring in IoT environments. Multiple channels like Wi-Fi, messaging services, and mobile applications ensure a reliable link between the sensors for data transmission. The major hardware components utilized here are sensors to measure blood pressure, heart rate, body temperature, and Raspberry Pi 3. The prototype is developed in a hand-held form. However, the latency for the Wi-Fi communication channel is comparatively high (125.95 second average). Afterward, Al-khafajiy et al. [55] introduced a health monitoring framework based on wearable sensors for older people that enables the patients to take healthcare facilities from their home environment. The sensors used in this framework are pulse, temperature, blood oxygen, and blood glucose sensors. All the components are connected to Arduino UNO, and the collected data are sent to the patient's mobile application through a Bluetooth module. Finally, the smartphone application is used as a gateway to send the data to the cloud server. The doctors can easily monitor the sensors data and patient's record through the developed monitoring platform. However, no data analytics tools are used for automated decisions.

In another work, Semwal et al. [56] presented a cost-effective and portable healthcare platform to ensure essential health services from remote locations. The sensory elements utilized in this system are ECG, pulse oximeter, body temperature, and blood pressure sensors. ATmega328P microcontroller and Bluetooth module are used for data collection and transmission respectively. LabVIEW tool is utilized for data visualization in the cloud server. The proposed prototype provides the offline accumulation of data from various sensors in low network connectivity that would be updated to cloud accessibility. However, no security measures are taken into consideration in this system. Afterward, Kumar et al. [57] introduced a smart healthcare monitoring system where patients and doctors can interact through a camera. The health parameters from input sensors (HR and temperature sensor) are sent to the processing module (Raspberry Pi) and displayed on LCD. The processed data are transmitted to the web server using an internet connection that assists physicians in monitoring the patient status in real-time. However, the developed system is linked to a limited number of sensors which were not enough to monitor a patient's complete status. Furthermore, Wan et al. [21] presented a wearable health monitoring framework that assists doctors in monitoring the patients in real-time through an IoT network. In this system, blood pressure, heartbeat, and body temperature sensors are used in the sensing node, and Arduino is used as an edge device. All the collected data are transmitted via Wi-Fi to the cloud server using the body area sensor network. The physicians can monitor the patients through their own devices like a laptop. However, the developed prototype did not mention any solutions for emergency cases.

Table 1 summarizes the developed health monitoring systems considering some properties like the used sensors and edge devices, the communication channel, the data visualization tools for ensuring the real-time monitoring, and comments of each developed system.

### B. SMARTPHONE SOLUTIONS FOR HEALTH MONITORING
The growing penetration of mobile phones and integrated sensors, and advanced communication technologies make it an appropriate infrastructure that allows continuous and virtual monitoring of patients' health. The built-in sensors in smartphones for health monitoring are a camera, accelerometer, gyroscope, proximity sensor, microphone, light sensor, and Global Positioning System (GPS) [23].

**TABLE 1.** Summary of the Smart Health Monitoring Systems Based on Wearable Devices

| Authors | Year | Sensors | Edge Devices | Communication Media/Protocol | Visualization | Comments |
|---|---|---|---|---|---|---|
| Islam et al. [20] | 2020 | Heartbeat and body temperature sensor | NodeMCU | Wi-Fi/ HTTP | ThingSpeak | The developed prototype is not adequately manufactured for application purposes. |
| Elango and Muniandi [50] | 2020 | Heartbeat sensor, body temperature sensor, and accelerometer | NodeMCU | Wi-Fi/HTTP, MQTT | ThingSpeak, LCD | No emergency services are mentioned in the proposed system. |
| Al-Sheikh and Ameen [51] | 2020 | Heart rate, ECG, SpO2, and body temperature sensor | Arduino, NodeMCU | Wi-Fi | Blynk | The system is not able to handle the emergencies of the patients. |
| Chigozirim et al. [52] | 2020 | Heartbeat and body temperature sensor | ATmega328P, NodeMCU | Wi-Fi | ThingSpeak, LCD | The developed system has not been fully fabricated for real-time tests. |
| Mohapatra et al. [53] | 2019 | Heartbeat and body temperature sensor | Arduino, NodeMCU | Wi-Fi/ HTTP | Adafruit | No security concerns are mentioned here for the transmitted data. |
| Swaroop et al. [54] | 2019 | Heartbeat, blood pressure, and body temperature sensors | Raspberry Pi | Wi-Fi/ MQTT | Web application | The latency for the Wi-Fi communication channel is comparatively high. |
| Al-Khafajiy et al. [55] | 2019 | Pulse, temperature, blood oxygen, and blood glucose sensors | Arduino | Bluetooth HC-06 | Web application | No data analytics tools are used for automated decisions. |
| Semwal et al. [56] | 2019 | ECG, pulse oximeter, body temperature, and blood pressure sensors | ATmega328P | Bluetooth HC-05 | LabVIEW | No security measures are taken into consideration in this system. |
| Kumar et al. [57] | 2018 | Heartbeat, and temperature sensor | Raspberry Pi | Wi-Fi/ HTTP | Web application | A limited number of sensors are not quite enough to monitor the patient's complete status. |
| Wan et al. [21] | 2018 | Heartbeat, blood pressure, and body temperature sensors | Arduino | Wi-Fi/ HTTP | Web application, LCD | The developed prototype did not mention any solutions for emergency cases. |

The major health parameters that can be monitored through smartphone sensors are heart rate and variability, blood pressure (BP), oxygen levels (SpO2), and respiratory rate. They are used to identify skin, eye, ear diseases. As almost all people are now using a smartphone, it has become a great choice to research smartphone applications that ensure portability and reduce the additional cost of the developed systems [58]–[60]. The systems that are designed for health monitoring using data collected from smartphone sensors are discussed here.

Zhang et al. [24] proposed a framework for blood glucose monitoring using PhotoPlethysmoGram (PPG) data in the form of a video from a smartphone. The blood glucose level is collected from patients through smartphones, and the collected data is processed in the cloud. Lastly, the features are extracted using Gaussian Fitting and classified into normal, warning, and borderline based on blood glucose range using machine learning algorithms. The accuracy for blood glucose level estimation achieved from the developed system is 81.49%, which can be deemed lower than needed for a reliable glucose monitoring system. In another work, Nemcovaa et al. [61] developed a framework for monitoring the SpO2, BP, and heart rate (HR) utilizing a mobile phone. The rear camera and microphone of a smartphone are used as sensing elements in this system. The camera data is converted to PPG and used for heart rate and oxygen saturation estimation. At the same time, the blood pressure is estimated from PPG and phonocardiogram (PCG) recorded by the microphone. A smartphone application is developed to determine the feasible position of the data collection device for blood pressure estimation. However, the synchronization between PPG and PCG signals is not handled although there are different time bases. Afterward, the authors of [62] introduced a respiratory monitoring system using smartphone sensors based on imaging and the Fourier transform technique. The skin surface video data is captured in the presence of a flashlight using an embedded smartphone camera and Plethysmography data is collected using developed hardware. The collected data is transferred to a PPG signal, and the processed data is analyzed using the concept of the discrete wavelet transform to estimate respiratory rate. The experimental results depicted that the system obtained an average accuracy of 97.8% and an average error of 2.2%. However, the system did not consider temperature and skin colour as well as the condition effect.

Recently, Tabei et al. [25] presented a framework for monitoring blood pressure using smartphones' cameras. The data for the proposed scheme is collected from the user's finger index through a smartphone camera. Filtering and peak detection techniques are used to minimize the motion and noise from the collected PPG signals. The estimation of blood pressure is done with the use of a linear regression algorithm. It is revealed from the experimental results that the system obtained mean absolute error, standard deviation, and correlation parameters of 2.10, 1.96, and 0.90, respectively. However, the derivation of pulse transit time from two sides of the arterial is not mentioned. In another research, Dey et al. [63] proposed a cuff-less blood pressure measurement system using a heart rate sensor embedded in a smartphone. The sliding window technique is used to convert the collected PPG signals to 15 s epochs. Approximately 233 features are derived from the raw signals from a PPG pulse in the domain of time and frequency. Finally, the blood pressure estimation is conducted using the Lasso regression technique. A smartphone application is developed to monitor the psychological signals in real-time. The developed system can calculate the 95% confidence interval of the BP of the patient. However, the results do not satisfy the precision in terms of standard value. Using the built-in accelerometer and camera of a smartphone, Wang et al. [64] introduced a blood pressure monitoring device named Seismo to interpret the vibration generated by heartbeat and finger pulses. The blood ejection time is measured from the seismocardiography signal through the accelerometer, and the fingertip is used to calculate the arrival time from PPG data using a camera module. The embedded speaker synchronized the data from the accelerometer and camera. The developed system obtained a Pearson correlation coefficient between 0.20 and 0.77 for the volunteers. However, the prototype is not able to monitor blood pressure continuously.

Further, a heart rate monitoring framework based on photoplethysmographic data from the smartphone is proposed [65]. The data is collected using the visible light reflected mode of PPG using a built-in smartphone camera from the user's index fingertip. The data from smartphone storage is transferred to the processing device using Bluetooth communication. Among the three channels of the PPG signal, the red channel is utilized for heart rate estimation in this system. The proposed scheme appraised an accuracy of 99.7%, and the found absolute error is within the range of 0.04–0.3 beats/min. However, the duration of the video is relatively low. In another study, Lomaliza and Park [66] developed a reliable and accurate system for HR monitoring using the camera images of a mobile phone of the fingertip. In this system, the signal is extracted using the concept of Region of Interest, and the noise from the raw data is eliminated through the adaptive threshold method. The developed scheme is adopted in any level of smartphone. The experimental finding depicted that the developed framework estimated heart rate in real-time having less than 5% error rate. However, the proposed system ignores the effects of different camera modules of different smartphones. Furthermore, Qayyum et al. [67] demonstrated a vital sign monitoring system using video from a smartphone camera. In this framework, the collected data is pre-processed to reduce the noise from the raw signals using the colour distortion filtering technique. The inter-beat interval is used to detect heart rate variability, and the breathing rate (BR), heart rate, and SpO2 are calculated from PPG signals. The developed system obtained a mean absolute percentage error of 2.965 from the experimental findings. However, no real-time prototype is shown in this study.

Table 2 summarizes the developed health monitoring systems considering some properties like the monitored signs, the used sensors, the smartphone model, the number of subjects, the video length, sampling rate, and comments of each developed system.

## III. ARTIFICIAL INTELLIGENCE AND MACHINE LEARNING IN IOHT

The Internet of Health Things comprises various interlinked devices that can share and handle data to enhance patient health. It has become a fast-growing area with numerous investments associated with the development and use of IoT [68], [69]. Statistics from the McKinsey study depicts that IoHT will have a financial impact of $11.1 trillion in a year by 2025 [70]. Machine learning has become a major tool in the arsenal of artificial intelligence techniques used in healthcare. It enables IoT devices with outstanding capabilities for information inference, data analytics, and intelligence. Machine learning has become indeed a powerful and effective solution for various IoHT technology contexts, from big-data cloud computing to smart sensors [71]–[73]. An overall system architecture for disease diagnosis using machine learning algorithms in the IoHT environment is shown in Figure 3. The used data in these frameworks are from benchmark datasets or real-time sensor data sent to the cloud for processing. Afterward, the data are preprocessed, and necessary features are extracted to fit in the machine learning techniques. Finally, the decision is transferred to the concerned person to take proper action. The significant developments of machine learning-based IoHT solutions are demonstrated in this section. We have described some major disease solutions using machine learning in the IoHT platform that are becoming significant threats for human-being in recent times.

### A. NOVEL CORONAVIRUS (COVID-19)

Novel coronavirus (COVID-19) has become a public health crisis due to this virus's communicable nature in recent times. This is an ongoing pandemic, and all the sectors of the whole world are fighting to recover from this ailment. The statistic shows that approximately 98 million cases have already found, and the death cases are about 2 million worldwide [74]. Numerous works have been conducted to reduce the severity of this disease using modern technologies [75]–[77]. Although several works have already been done, we described the frameworks that used machine learning algorithms to diagnose COVID-19 in IoT environments.

**TABLE 2.** Summary of the Smart Health Monitoring Systems Based on Smartphone

| Authors | Year | Signs | Sensors | Smartphone | Number of participants | Video length | Sampling rate | Comments |
|---|---|---|---|---|---|---|---|---|
| Zhang et al. [24] | 2020 | Glucose level | Camera | iPhone 6s Plus | 80 | 60 s | 30 Hz | The performance of the developed system is relatively low. |
| Nemcovaa et al. [61] | 2020 | HR, BP, SpO2 | The rear camera, microphone | Honor 7 Lite, Apple iPhone SE, Lenovo Vibe S1 | 22 (13 F, 9 M) | 20 s | 30 Hz | The synchronization between PPG and PCG signals is not handled, although there are different time bases. |
| Alafeef and Fraiwan [62] | 2020 | Respiratory rate | Camera | Samsung Galaxy S6 | 15 | 30 s | 30.30 fps | The system did not consider temperature and skin colour as well as the condition. |
| Tabei et al. [25] | 2020 | BP | Camera | iPhone X | 6 | 120 s | 30 fps | The derivation of pulse transit time from two sides of the arterial is not mentioned. |
| Dey et al. [63] | 2018 | BP | Heart rate sensor | Samsung Galaxy S6 | 205 (90 M, 115 F) | 15 min | 125 Hz | The results do not satisfy the precision in terms of standard value. |
| Wang et al. [64] | 2017 | BP | Rear camera, accelerometer | N/M | 9 | N/M | N/M | The prototype is not able to monitor blood pressure continuously. |
| Alafeef [65] | 2017 | HR | Camera | Samsung Galaxy | 19 | Few seconds | 30.30 fps | The duration of the video is relatively low. |
| Lomaliza And Park [66] | 2017 | HR | Rear camera | Pantech Vega Racer 3, Samsung Galaxy S2, LG G3, HTC Desire HD | 5 (3 M, 2 F) | 10 s | 10 fps | The proposed system ignores the effects of different camera modules of different smartphones. |
| Qayyum et al. [67] | 2017 | HR, HR variability, BR, SpO2 | Camera | iPhone 4 | 20 (10 M, 10 F) | 60 s | 30 fps | No real-time prototype is shown in this study. |

M = Male, F = Female, N/M = Not Properly Mentioned.

Very recently, Le et al. [26] developed IoT based system to diagnose COVID-19 using the concept of convolution neural network (CNN) and support vector machine (SVM). In this framework, the data is retrieved from the patients utilizing IoT sensors and transferred to cloud storage through 5G networks. In addition, the CXR dataset [78] is utilized to conduct an experiment. To reduce noises from the raw images, Gaussian filtering is utilized. The depth-wise separable CNN extracted the features from the pre-processed samples, and SVM categorized the extracted features to detect COVID-19. The proposed system appraised an accuracy of 98.54% for binary class and 99.06% for the multiclass scenario. However, no monitoring system from the doctor's end is developed here. Afterwards, Ramallo-González et al. [79] introduced an IoT platform named CIoTVID for the detection of coronavirus. The scheme includes various levels of sensorization, which can handle and evaluate the data that assists in making a decision. The data collection layer collects various symptoms like voice signals, oxygen saturation, respiration rate from the patients. Mel Frequency Cepstral Coefficients (MFCC) transferred the raw voice signals to spectrogram as an image format. CNN architecture extracted the features and classified them in this system. The use case analysis found that the system appraised an accuracy of 66.67% in the testing phase. However, the outcome is relatively low for real-time use. In another research, Ahmed et al. [80] proposed a deep learning-based framework to diagnose COVID-19 in an IoT environment. A combined architecture like Faster-RCNN with ResNet-101 is utilized to diagnose the coronavirus cases from chest X-ray samples. The used data are retrieved from various open-access data sources where the COVID-19 cases are about 4000, and the negative cases are 7000. The data from the medical sensors are directly sent to cloud storage using Wi-Fi communication, where the proposed architecture is trained and finally diagnoses the positive cases. The radiologist can monitor the outcome through the internet. It is evident from the experiments that the developed system obtained an accuracy of 98%. However, no usability study is mentioned in this system.

Otoom et al. [81] introduced a scheme using machine learning techniques in an IoT environment to detect and monitor coronavirus-infected patients. The real-time symptom data from the patient's end are retrieved utilizing IoT devices and transferred to the cloud server for storage. Benchmark data called COVID-19 Open Research Dataset (CORD-19) [82] is utilized for the analysis in addition to real-time data. The collected data are analyzed using machine learning classifiers to detect coronavirus infections. Among the eight classifiers, neural networks and k-nearest neighbors performed the best, and the accuracy was 92.89%. The system notified the medical experts of the suspected cases for further clinical treatment. However, the performance of the developed system is relatively low for practical uses. Further, ElRashidy et al. [83] demonstrated a deep learning architecture based on end-to-end nature to diagnose and monitor coronavirus-infected patients. In the proposed system, the patients are monitored through wearable sensors and smartphone app; a fog network is used to handle the data storage and transmission issues, and finally, a CNN architecture with transfer learning diagnosed the COVID-19 patients from X-ray samples. The experimental data is collected from two publicly available datasets [78], [84], and wearable sensors. It is found from the experiments that the developed scheme obtained accuracy and specificity of 97.95% and 98.85%, respectively. The physicians monitored the patients in real-time and guided the individuals properly. However, energy consumption and storage issues are still a challenge for this system. In another research, Karmore et al. [27] developed humanoid software for the diagnosis of coronavirus in IoT networks that can identify whether an individual is infected with this ailment or not. The robotic system used IR sensors and a camera module for navigation, and the E-Health sensor kit and chest X-ray scans are utilized for diagnosis. The developed humanoid robot used NodeMCU, Raspberry Pi, temperature sensor, ECG sensor as hardware components. Three pre-trained architectures like InceptionV3, ResNet50, and Inception ResNetV2 diagnosed positive cases from X-ray samples. The average accuracy found from the proposed system is 97.95%. However, the security issues during the data transfer are not depicted here.

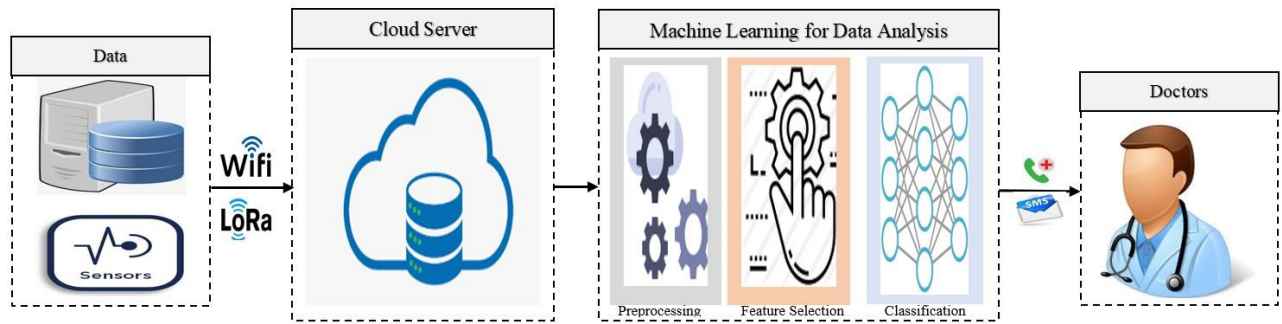

**FIGURE 3.** An overall system architecture of machine learning-based framework for disease diagnosis in IoHT environment.

Furthermore, Cacovean et al. [28] introduced an IoT-based framework for COVID-19 detection where machine learning techniques are utilized for diagnosis. The data from the participants are collected using wearable devices like GPS, temperature, and heart rate sensors. The retrieved data are sent to the oracle cloud server for processing through the Bluetooth module. Random Forest obtained the best outcome from the experiments among the three classifiers, and the accuracy value is 73%. The prediction outcomes are directly sent to doctors and patients' guardians to take further steps for proper treatment. However, the system achieved comparatively low performance. Afterward, Kumar et al. [85] presented a system to monitor the COVID19 patients using sensor and IoT technology. The real-time data are retrieved from the participants using IoT sensors and fed into the Bayesian network for preprocessing. The IoT devices are configured and accessible using wireless sensors to send the data to the patient's repository. The data are trained with SVM and predict the coronavirus cases from the test samples. The scheme appraised an accuracy of 87.23% and 86% for recovery and prediction, respectively, using SVM. In addition, the K-means algorithm estimated the spread as well as recovery rate. However, the accuracy rate of the system is not up to the mark.

Table 3 briefly discusses COVID-19 detection systems highlighting some features such as the used datasets, the used techniques for detection, the accuracy as a performance metric, and comments of each reviewed system in IoHT environments.

### B. HEART DISEASE

Heart disease has become a very crucial and acute ailment for every aged people, especially for adults. An estimation shows that heart disease is responsible for approximately 30% (18 million individual) deaths among all death cases per year [86], [87]. Hence, the researchers are focusing on the development decision support system in the smart healthcare environment to reduce the severity of heart disease. The significant developments of heart disease diagnosis using machine learning in the IoT environment are demonstrated here.

Recently, the author of [88] developed a patient monitoring system for heart patients in an IoT environment where the data from the patients are analyzed using a modified Deep Learning Modified Neural Network (DLMNN). The body-worn sensors collected data from the patients and securely sent them to the cloud for further processing. In addition, the proposed system used the Hungarian heart disease (HD) dataset [89]; benchmark data for heart disease classification to detect the presence of abnormality. An alert message is delivered to the doctors while any abnormality is found. However, the developed scheme obtained comparatively low performance in the case of small data size. In another work, Ali et al. [29] proposed a smart healthcare monitoring framework for heart disease infected patients using the concept of ensemble learning and feature fusion. The extracted features from sensor data and patient history are merged through the feature fusion technique in this system. The information gain method eliminated the unnecessary and redundant features, selected the most appropriate features responsible for the disease. A semantic web rule language is introduced that recommends the activities of the infected patients automatically. Lastly, the LogitBoost technique, an ensemble learning classifier, is used to predict heart disease and obtained an accuracy of 98.5% from the experiments. However, the developed system used traditional techniques for feature selection, reduction, and classification. Afterward, Deperlioglu et al. [30] introduced a framework for heart disease diagnosis using an autoencoder network in an IoHT environment. The developed system comprises a cloud environment where beacons are used for data sharing and a central system to synchronize the cloud and devices' communication and machine learning architecture. Two heart sounds datasets named PASCAL BTraining [90], and Physiobank-PhysioNet A-Training [91] are used in this system. The developed system obtained accuracy, sensitivity, and specificity of 100% for the PASCAL dataset. In addition, an accuracy of 96.03%, 91.91%, and 90.11% are achieved for healthy heart sounds, extrasystole, and murmur, respectively, from 479 real-time participants. However, no voice command facility is available in this study to ensure less physical interaction.

**TABLE 3.** Summary of COVID-19 Detection Frameworks in IoHT Environment

| Authors | Year | Data | Techniques | Accuracy (%) | Comments |
|---|---|---|---|---|---|
| Le et al. [26] | 2021 | CXR dataset and real-time sensor data | CNN, SVM | 99.06 | No monitoring system from the doctor's end is developed. |
| Ramallo-González et al. [79] | 2021 | Real-time sensor data | CNN | 66.67 | The outcome is relatively low for real-time use. |
| Ahmed et al. [80] | 2020 | Various public datasets | Faster-RCNN, ResNet-101 | 98 | No usability study is mentioned. |
| Otoom et al. [81] | 2020 | CORD-19 dataset, and real-time sensor data | Eight machine learning techniques | 92.89 | The performance of the developed system is relatively low for practical uses. |
| El-Rashidy et al. [83] | 2020 | Two open-access datasets and real-time sensor data | ResNet-50 | 97.95 | Energy consumption and storage issues are still a challenge for this system. |
| Karmore et al. [27] | 2020 | Real-time sensor data | InceptionV3, ResNet50, Inception ResNetV2 | 97.95 | The security issues during the data transfer are not depicted. |
| Cacovean et al. [28] | 2020 | Real-time sensor data | Random Forest, Naïve Bayes, SVM | 73 | The system achieved comparatively low performance. |
| Kumar et al. [85] | 2020 | Real-time sensor data | K-means algorithm, SVM, Bayesian classifier | 86 | The accuracy rate of the system is not up to the mark. |

In another research, the authors of [31] presented a machine learning-based heart disease diagnosis system in the IoMT cloud environment using modified salp swarm optimization (MSSO) and an adaptive neuro-fuzzy inference system (ANFIS). The data from the IoMT sensors, as well as UCI [89] and Framingham database [92], are used to diagnose the presence of heart disease. The MESO technique optimized the dataset's attributes to find the best features, and the ANFIS trained the most appropriate features and diagnosed the disease. The experimental results found that the system achieved accuracy, AUC, and precision of 99.45%, 99%, and 96.54%, respectively, using the datasets, yet no results are presented for real-time data. The authors of [93] proposed an IoT-based system using Modified Deep Convolutional Neural Network (MDCNN) to predict heart disease. The data (BP and ECG) used for this study are collected from smartwatches and heart monitor devices attached to the patient's body. In addition, some open-access databases like UCI [89], Public Health, and Framingham [92] are also used to train the network. In this framework, Long-range (LoRa) communication protocols, LoRa cloud, and servers are used to ensure the real-time monitoring of the patients. The developed framework categorized the sensors' data into two classes (normal and abnormal) and obtained an accuracy of 98.2% from the experiments. However, no wearable prototype is mentioned here. Further, Tuli et al. [94] developed a smart healthcare framework named HealthFog to diagnose heart disease using ensemble learning in IoT and Fog computing environments. To evaluate the performance of the developed system with respect to energy consumption, accuracy, latency, and execution time, FogBus (Fog-based cloud environment) is utilized. The FogBus is comprised of the worker node, the cloud data center, and the broker node. The Bagging classifier categorized the data collected from sensors and benchmark datasets [89] of heart disease. The developed prototype achieved an accuracy of 89% from the experiments for the test cases. As the machine learning architecture is trained in each worker node of every fog node, the time consumption becomes comparatively high.

In another study, Nguyen et al. [95] introduced a scheme for diagnosing heart disease using machine learning in an IoT environment. ECG devices collect the data from patients and send them to cloud storage through Wi-Fi. Wavelet-based Kernel Principal Component Analysis (wkPCA) technique pre-processed the raw data and extracted the most relevant classification features. The extracted features are fed into a neural network that diagnoses the heart disease based on the input data. The developed system achieved an accuracy of 98.03%. However, no notification system to alert physicians is developed yet. Furthermore, the authors of [96] demonstrated an IoT-based framework to diagnose heart disease using cloud storage and machine learning algorithms. The data is collected from the human body using medical IoT sensors, and benchmark datasets [97] from UCI are also used for the experiments. All the data are stored in a cloud database, and machine learning techniques are applied to the cloud database to predict the presence of heart disease. Among four classifiers, J48 performed the best and obtained accuracy, precision, recall, and F1-Score of 91.48%, 91.50%, 91.50%, and 91.50%, respectively. However, no real-time study is illustrated in this system.

Table 4 summarizes heart disease detection systems considering some features such as the used datasets, the used algorithms for detection, the accuracy as an evaluation metric, and comments of each developed system in the IoHT environment.

*C. DIABETES*

Diabetes is another life-threatening disease for humankind that results in many deaths per year. An estimation shows that almost 463 million individuals had diabetes in 2019, and the numbers are expected to grow to 578 million and 700 million by 2030 and 2045, respectively [98]. As this ailment is rising rapidly, early diagnosis of diabetes is necessary for the sake of people. Various studies are conducted to diagnose diabetes early, utilizing artificial intelligence, IoT, and Big data [99]–[101]. The works that are developed recently for diabetes detection are illustrated in this section.

In recent times, Rghioui et al. [32] developed a framework to monitor and predict diabetic patients using machine learning techniques in IoT networks. In this system, the glucometer is connected to NodeMCU to record the data from patients seamlessly. A vast amount of collected data are sent to the cloud database using IoT platform and processed using machine learning algorithms, and the decisions are sent to doctors for further treatment. The dataset comprises five features with 12612 records. Among the four algorithms, Random Forest achieved an accuracy of 96.05% from the experiments. In another research, Allugunti et al. [102] proposed a diabetes prediction framework using the concept of IoT and a decision tree to monitor the infected patients in real-time. The data is collected from IoT sensors and contained eight attributes and 15,000 realities. The best features from the dataset are selected using the concept of entropy measurement. The experimental outcomes show that the developed framework obtained accuracy and an error rate of 96.43% and 5.37%, respectively, in prediction. However, the detailed data collection procedure is not mentioned here. Afterward, Efat et al. [103] demonstrated a health monitoring system focusing on diabetic patients that can monitor the level of sugar, sleep time, food intake, and pulse rate. The data from the patient's side is continuously sent to a neural network using wearable sensors through Bluetooth and the developed architecture categorized the data based on the severity of diabetes cases. An alert message/call is sent to the patient's guardians and caregivers in an emergency. The developed scheme appraised an accuracy of 84.29% from 25 diabetes patients' data. A web portal is also developed to monitor the patients' health status continuously. However, the performance is relatively low for practical use.

In another work, a diabetes monitoring and prediction framework is proposed in [104] utilizing IoT and machine learning techniques. The system used a blood glucose meter, Arduino, and GSM modem as hardware components. The experimental data are retrieved using the glucose meter using an edge device like Arduino and processed in the microcontroller. The decision of the processed data is automatically sent to mediators through a GSM modem. The author found the Random Tree classifier to provide the highest accuracy, and lowest training time among the four used algorithms, and the value are 97.87% and 0.03 seconds, respectively. However, the glucose sensor and Arduino could not be operated at the same time. Further, Rghioui et al. [105] introduced an intelligent framework for diabetes-infected patients monitoring using machine learning architectures in IoT networks to monitor physical activity, glucose level, and body temperature. For data collection purposes, a glucometer, temperature sensor, and motion sensor are used at the patient's end. The collected data are transferred to the database station using a smartphone through 5G networks. The patient's records are classified using six classification algorithms, and the minimal sequential optimization (SMO) obtained the best accuracy of 99.66%. Whenever any abnormality is found, a notification text is sent to the doctors to take proper steps for treatment. However, the latency is comparatively high in this system. Afterward, Godi et al. [106] developed a healthcare monitoring framework to diagnose and monitor disease using machine learning modalities through IoT networks. Various wearable devices are utilized to retrieve data from patients from different scenarios like homes and hospitals. In addition, a diabetes dataset from the Kaggle repository [107] is used for experiments. Machine learning techniques classified the data based on the presence of abnormality. Among four classifiers, SVM achieved an accuracy of 80.51%, precision of 76%, recall of 65%, and F1-Score of 70% for the positive diabetes class. The predicted results are shared with the physicians, mediators, and patient's caregivers. However, the performance is relatively low for real-time implementation.

**TABLE 4.** Summary of Heart Disease Detection Systems in IoHT Environment

| Authors | Year | Data | Techniques | Accuracy (%) | Comments |
|---|---|---|---|---|---|
| Sarmah [88] | 2020 | Hungarian HD dataset and real-time sensor data | DLMNN | 96.8 | The developed scheme obtained comparatively low performance in the case of small data size. |
| Ali et al. [29] | 2020 | Cleveland dataset, and real-time sensor data | Feature fusion and ensemble learning (LogitBoost) | 98.5 | The developed system used traditional techniques for feature selection, reduction, and classification. |
| Deperlioglu et al. [30] | 2020 | PASCAL and PhysioNet dataset, and real-time sensor data | Autoencoder neural network | 96.03 | No voice command facility is available in this study to ensure less physical interaction. |
| Khan and Algarni [31] | 2020 | Hungarian and Framingham dataset, and real-time sensor data | MSSO-ANFIS | 99.45 | No real-time study is shown in this system. |
| Khan [93] | 2020 | Cleveland database and real-time sensor data | MDCNN | 98.02 | No wearable prototype is mentioned. |
| Tuli et al. [94] | 2020 | Cleveland dataset, and real-time sensor data | Ensemble learning(Bagging) | 89 | The computational time is relatively high. |
| Nguyen et al. [95] | 2020 | Real-time sensor data | wkPCA, BNN | 98.03 | No notification system to alert physicians is developed. |
| Ganesan and Sivakumar [96] | 2019 | Statlog dataset and real-time sensor data | J48, logistic regression, multilayer perceptron, SVM | 91.48 | No real-time study is illustrated in this system. |

To predict diabetes mellitus, the authors of [33] introduced a novel framework using machine learning in the IoT environment. The glucose sensors collected the blood sugar data from the patients, normalized it into the proper format, and transferred it to the storage device using HTTP and MQTT protocol. The system also used a benchmark dataset for the experiment. Two machine learning classifiers named SVM and KNN are used for diabetes prediction, where SVM achieved accuracy and F1-Score of 90% and 89%, respectively. However, real-time cases are not found there. Furthermore, Kaur et al. [108] introduced a framework named CI-DPF to predict diabetes in a cloud-based IoT environment. The blood glucose level from the patients is collected using smart sensors and sent to the cloud environment for storage and further processing through IoT devices. The proposed system also used benchmark data named Pima Indians Diabetes dataset [107] for the experiment. Ensemble learning is used to diagnose diabetes from the patient's records, and it is found that the ensemble of decision tree and neural network obtained accuracy, sensitivity, and specificity of 94.5%, 79.5%, and 83.12%, respectively. However, real-life clinical tests have not been conducted here.

Table 5 briefly discusses diabetes detection frameworks highlighting some properties such as the used datasets, the used algorithms for detection, the accuracy as a performance metric, and comments of each reviewed system in the IoHT environment.

## IV. AMBIENT ASSISTED LIVING

Ambient Assisted Living involves combining sensors and actuators in an IoT environment to communicate and provide enhanced lifestyle and human-independent care for older adults. With 87% of older adults preferring living at their own homes over senior homes [109], there are two target environments for AAL, senior care homes and older adults' private homes. We first go in-depth into the studies carried out to define the needs of the seniors, then discuss the implementations both with and without a robotic social agent.

### A. USER NEEDS STUDIES

Several studies are directed towards studying the needs of older adults. Such requirements can be categorized into physical assistance, emotional support, reminders, or social support. Bedaf et al. [38] performed a user study with different stakeholders, including 11 formal caregivers, seven informal caregivers, and ten older adults as part of the ACCOMPANY project. After the users interacted with a robotic system (Care-O-Bot 3) in a fetch-and-carry scenario and a scenario where the robot reminded them to drink water, the authored received several suggestions on the functionalities they need. As such, the authors concluded that a social robot for elderly care needs to have advanced speech interaction capabilities, fetch and carry various objects, detect dangerous situations, alert the caregivers, and be adaptable to individual user needs. Likewise, the HomeMate project [39] defined five main scenarios that would benefit users the most: fetch-and-carry, infotainment (music and movies), gaming services, video chatting, and reminders for various events. After implementing a physical prototype and its testing with older adults, the authors emphasized the importance of natural interaction, specifically through speech. Other studies, such as [110], highlight the importance of combining both AAL sensor technologies with a robot for task achievement as well as social companionship. Similar conclusions were made in [111], where the authors highlighted how 50% of older adults requested efficient speech interaction and added that the inclusion of a robotic platform provides multiple benefits. Syed et al. [36] highlight that movement should be monitored for older adults, citing falls as one of the major causes of death in the senior community. Other studies that highlighted the needs of older adults include [37], [41], [112], [113]. The requirements can be summarized as:

- Provide a natural means of interaction that require minimum to no learning by the older adult.
- Remind the users of medications, appointments, and events.
- Provide infotainment services such as music, movies, and cognitive games.
- Real-time monitoring of health vitals and detection of emergencies.
- Include a robotic platform for task achievement as well as social companionship.

**TABLE 5.** Summary of Diabetes Detection Frameworks in IoHT Environment

| Authors | Year | Data | Techniques | Accuracy (%) | Comments |
|---|---|---|---|---|---|
| Rghioui et al. [32] | 2021 | Real-time sensor data | Naive Bayes, Random Forest, OneR, SMO | 96.05 | No user prototype is shown in this system. |
| Allugunti et al. [102] | 2020 | Real-time sensor data | Decision Tree | 96.43 | The detailed data collection procedure is not mentioned. |
| Efat et al. [103] | 2020 | Real-time sensor data | Neural network | 84.29 | The performance is relatively low for practical use. |
| Rghioui et al. [104] | 2020 | Real-time sensor data | Six machine learning algorithms | 97.87 | The glucose sensor and Arduino could not be operated at the same time. |
| Rghioui et al. [105] | 2020 | Real-time sensor data | Six classification algorithms | 99.66 | The latency is comparatively high in this system. |
| Godi et al. [106] | 2020 | A benchmark data from Kaggle and real-time sensor data | Decision tree, regression, K-NN, SVM | 80.51 | The performance is relatively low for real-time implementation. |
| Ponmalar and Vijayalakshmi [33] | 2019 | Real-time sensor data | K-NN, SVM | 90 | Real-time cases are not found in this study. |
| Kaur et al. [108] | 2018 | Pima Indians Diabetes dataset and real-time sensor data | Ensemble of decision tree and neural network | 94.5 | Real-life clinical tests have not been conducted. |

In the remaining section, the implementation of various platforms for AAL is discussed. First, we showcase research work done with IoMT environments without robotic agents, followed by studies that involved social robots as a central component of the system.

### B. SMART HOMES

Komai et al. [34] present a system to monitor the activity of multiple seniors simultaneously based on Bluetooth Low Energy (BLE) with a beacon in the user's name card and the Received Signal Strength Indicator (RSSI). Likewise, [114] uses BLE and RSSI to create a low-cost indoor-localization method to track and estimate the user's room. They propose the method as a low-cost system to detect older adults' activity and early signs of frailty using a Random Forest classifier. Although both systems are proposed based on low-cost solutions for indoor localization, they can only achieve room accuracy.

Marques et al. [109] present an indoor environmental monitoring system that measures room temperature, relative humidity, $CO$, $CO_2$, light detection, and transmitting the messages through XBee through Zigbee networking protocol. The system proves to be a modular and cheap solution for indoor air quality monitoring. The work is extended in [35], where a robotic platform is integrated equipped with a gas sensor to detect levels of liquefied petroleum gas (LPG), isobutane, and propane, which can lead to explosions when they reach specific levels. Based on the famous Turtlebot platform, the robot can use that sends a notification to the user through the Facebook social platform. The system provides a safe way for monitoring gas levels but allows for little to no control by the user. Diraco et al. [115] propose a sensory system based on radars to monitor heart and respiration rates of older adults without contact (from a distance), achieving 95% and 91% accuracy, respectively. The authors also utilized the radar for fall detection, resulting in a sensitivity of 97% and specificity of 90%. Nevertheless, ultra-wideband radio signals can measure a few vital signs and need a more versatile system to integrate more sensors.

The authors of [36] propose a framework for monitoring physical activities and utilizing machine learning algorithms for more accurate and faster predictions and decisions. They build on the mHealth framework, initially proposed in [116] to collect data from multiple sensors and combine them to predict 12 different physical activities using a multinomial Naïve Bayes classifier, achieving an accuracy of 97.1% on the mHealth dataset. Although the framework sounds promising, it is yet to be tested in a real-world environment.

A mobile application called InfoSage is offered by Quintana et al. [22] to connect older adults to their formal and informal caregivers, centred around the older adult as the keystone user. The solution focuses on dementia patients and offers a tracker and reminders for appointments and medications. The system offers capabilities to exchange messages between family members (informal caregivers), doctors, and older adults and share information. The authors also perform several user studies on the acceptance and usability of their platform.

Stavrotheodoros et al. [37] propose the IN LIFE platform, a cloud-based solution that combines various sensors focusing on personalization and easy installation for cognitively impaired older adults. The system is capable of monitoring user activities through unobtrusive sensors and multilayered architecture. The system is comprised of 3 layers, a perception layer with the various sensors for data collection, a gateway layer that combines the data and transmits them to the final layer (cloud layer) using MQTT protocol. The cloud layer stores and analyzes the data. The work is extended in [117] by establishing the system and using motion sensors and door sensors to identify user habits and a panic button for older adults when there is an emergency.

The authors of [118] propose an ambient assisted living system that utilizes fog computing. A system is put in place that incorporates radar sensors to detect daily activities and implements an algorithm to detect whether the patient is suffering from a neurological disease attack, if they are idle, or if no patient is detected. Utilizing the fog layer in the system leads to minimal response delay and energy consumption coupled with more bandwidth efficiency and overall performance.

Several of the proposed solutions provide promising solutions for AAL, all aiming at better and healthier living. Nevertheless, the area still requires much work, specifically applying the proposed systems in real-world environments for elaborate testing and user feedback.

### C. SOCIAL ROBOTS

Previous studies did not harvest the power of robotics to provide more functionalities for older adults. The following discussion tackles studies that involve social robots, a summary of the most prominent works is provided in Table 6. Portugal et al. offer SocialRobot in [40], a modular robotic platform with independent layers. SocialRobot can adapt to user's preferences and includes human-robot interaction (HRI), emotion and facial recognition, and speech interaction. The robot is based on ROS (Robotic Operating System) and uses a sequential database (MySQL) to store data on each for more personalized interactions, which the authors called SoCoNet. The work was extended in [120] and [111], where the authors studied the effect of including environmental context on the decision process and tested SocialRobot in an elderly care home for a week. Besides navigation and storing data on its users, SocialRobot can also recognize faces and user's emotions. After testing SocialRobot in an elderly care home in the Netherlands, the authors concluded that considering the current context improves the accuracy of predictions. They also noted that although the robot was generally deemed acceptable by the older adults, it should incorporate more effective speech interaction and anthropomorphism (arms) for better functionality, which it is not equipped with. SocialRobot covers several requirements of the older adult community. However, it has been shown that integrating social robots in an IoT environment with sensors that track older adults provides much greater promise. Moreover, the caregivers and older adults requested functionalities such as playing music, movies, games, and memory training activities.

The HomeMate project [39] involved studies defining older adults' requirements and three iterations of creating a robotic platform. The HomeMate robot can play movies and music, link older adults with their family members and friends through video chat. Besides that, it was also able to communicate through both touch and voice interaction and schedule reminders for different events, thus tackling the missing functionalities from the SocialRobot project. However, the users still requested more natural and intuitive speech interactions, possibly through a more intelligent natural language understanding module. Moreover, the robot was also missing a link to smart devices for monitoring environmental conditions and users' health.

While both HomeMate and SocialRobot tackled senior home environments, Gross et al. proposed a robot companion for private homes called Sympartner [41]. The companion can provide reminders, health updates, daily routines, greet visitors at the apartment door, detect and identify objects and faces. Sympartner had autonomous navigation but did not have a manipulator to carry objects. HRI was done through a graphical user interface as well as through simple speech commands. The robot was deployed with 20 participants in their private homes for five days each in Germany. Although the overall feedback from the participants was positive, they requested speech understanding capabilities. Moreover, the robot was reported to have failed several times and required remote teleoperation by the researchers, which presents a privacy concern to its users.

The previous studies involved social robot implementations. However, none of them were integrated into smart home environments with other environmental and health monitoring sensors. The idea of including robots in the AAL environment is recommended by [42], [43],

**TABLE 6.** Summary of the Social Robots Implementations for Care of Older Adults

| Authors | Year | Sensors in IoT Environment | Robot Functionalities | HRI Methods | Comments |
|---|---|---|---|---|---|
| Nasr et al. [42] | 2020 | Fitbit Versa2 Smart Watch | Autonomous navigation | Speech interaction | Needs validation for acceptance and usability through testing it with older adults in a real-world environment. |
| Marques et al. [35] | 2019 | Gas sensor: LPG, iso-butane, propane | Autonomous navigation | Notifications on Facebook (no user input) | The system lacks HRI methods and provides no control for users over the system. The system needs a user interface with more functionalities needed for older adults. |
| Lee and Naguib [39] | 2019 | N/M | Autonomous navigation, fetch-and-carry, object handover, telepresence (video chat) | Speech interaction, gesture control | The system could only simple speech interaction and is missing a link to an IoT environment with smart sensors. |
| Loza-Matovelle et al. [44] | 2019 | Custom bracelet: HR, body temperature, angular acceleration, and heading | Tele-operation, obstacle avoidance, facial recognition, telepresence (video chat) | Speech Interaction and vision | Needs validation for acceptance and usability through testing it with older adults in a real-world environment. The robot needs autonomous navigation to eliminate privacy concerns. |
| Gomez-Donoso et al. [119] | 2019 | Cameras | Autonomous navigation, depth vision, object detection | Alarms | The system needs efficient means for user interaction and sensors/wearables for tracking body vitals. |
| Portugal et al. [40], [111], [120] | 2018 | N/M | Autonomous navigation, personalized interactions, facial recognition, emotion detection | Touch Screen | The robot has no speech interaction capabilities and is not linked to an IoT environment with smart sensors. |
| Do et al. [43] | 2017 | Smartwatch, IMU, Passive Infrared Sensor, microphones | Autonomous navigation, sound detection | Touch screen UI (mobile phone or tablet) | It needs validation for acceptance and usability through testing it with older adults. |

[121]. This can be thought of as an Internet of Robotic Things (IoRT) framework. Nasr et al. introduce a solution for AAL environments with robotics in [42]. The platform connects heterogeneous agents such as mobile robots, virtual assistants, and mobile phones with smart sensors and wearables. The system was designed with an emphasis on human-robot interaction and intuitive speech interaction with its users. The authors developed the platform with a MySQL database for storing reminders, a mobile phone to utilize Google Assistant's speech-to-text functionality, a Fitbit Versa2 smartwatch for HR monitoring, and a simulated robot in ROS and Gazebo. Two different protocols for data sharing are provided in the system, MQTT and REST API, to allow all kinds of smart devices and agents. The system is shown to respond to natural ways of giving commands and without the training of users. It provides its users with control over the robotic system by sending navigation commands and teleoperation through speech. Also, the authors emphasize the idea of providing the same functionality across all devices through utilizing a common NLU and Dialogue Management agent. While the system shows great promise in integrating heterogeneous systems in a modular way, it still needs to be tested in a real-world environment and requires a functional robot capable of at least autonomous navigation and moving objects.

Do et al. propose a similar approach to the problem through the robot-integrated smart home platform (RiSH) [43]. They utilize a Pioneer 3-DX platform based on ROS and are equipped with a microphone for acoustic detection. The robot is linked to a network with body sensors, including an inertial measurement unit (IMU), motion sensors, and a smartwatch to obtain ECG, SpO2, and respiration rate readings and a home sensor network which includes a passive infrared for binary motion detection and microphones for acoustic data. The authors present a system that is extensible and capable of leveraging smart home sensing capabilities. Furthermore, they conduct experiments with 12 older adults to detect human trajectories down to a 0.2 meter accuracy and recognize 37 different human activities and falls with accuracy up to 88% and 80%, respectively.

Another study that follows the integration of robots for AAL is provided by Loza-Matovelle et al. in [44]. The system is composed of a heterogeneous network of sensors both on a robotic platform and a bracelet that the user wears. The bracelet can measure heart rate and body temperature and angular acceleration, and heading (gyroscope). The accelerometer and gyroscope are used to detect falls of the older adult and send warnings to the family members. The robot is capable of localization and obstacle avoidance but needs to be teleoperated. In addition, it is capable of facial recognition to maintain contact with the user during interaction and allows for video conferencing with formal and informal caregivers (telepresence). The user is provided with the ability to interact with the robot and a hologram (called the interactive pyramid) through speech and a chatbot. The interactive pyramid's speech interface updates the user on the weather, time, reminders on medications and visits, as well as health recommendations. In essence, the system is like the works in [42], [43]. However, it uses teleoperating, which creates a privacy risk and lower autonomy of the system, and the system is yet to be tested in a more realistic environment.

Gomez-Donoso et al. [119] integrate a robotic system into an AAL environment equipped with cameras to detect dangerous situations. The authors found that the existing system does not detect dangers such as objects on the floor, knives (which change location), and dangers in occluded areas. Therefore, they added a Pepper robot based on ROS with an RGB-D camera to detect such dangers. The robot was capable of ground plane detection and clustering of pixels to find objects on the floor. Also, as the robot moves, it used an R-CNN to detect smaller sources of danger such as electrical outlets and knives. Pepper was also capable of autonomous navigation and detection of people who have fallen. The robot was linked to Wi-Fi and would send out an alarm whenever the older adult is in danger. The authors tested the robot in IoT environments that cover residential areas, clinical areas (nursing home), and an office. The robot is shown to increase the ability of the AAL to detect potentially harmful circumstances in different scenarios. Such a system is powerful but would benefit when integrated with other sensors that monitor the user's health vitals.

## V. SOFTWARE INTEGRATION ARCHITECTURES

Global business processes in IoMT demand information to be shared quickly and efficiently across many different software, tasks, and applications. Having discussed the general areas of research in IoMT, we find much common ground in approaching and the architectures utilized. To unify the approach and paving the road from research to implementing the systems in real-time environments, various architectures are provided in literature that attempts an inclusive solution for an IoMT environment that is modular and easily extensible in its functionalities. However, each architecture is tailored to a specific use case. This section gives an overview of the most promising architectures and frameworks proposed in the area and an in-depth analysis of their advantages and shortcomings.

Petrovic et al. [122] popped the following question: Why´ recreate devices, sensors, and systems while we can use off-the-shelf solutions and augment their capabilities to provide a far more useful system. Therefore, the authors use commercial off-the-shelf (COTS) smart devices (COTS component) and combine them through an Interactivity component that provides an interface to retrieve data from the COTS devices and performs pre-processing stages. This layer can also detect some patterns and issue warnings in danger to a patient's health. The third and last component of the system is the Cloud component which synchronizes the COTS devices and takes care of big data analytics. While this solution offers faster solutions and lower costs, it is just mentioned as an idea, and no validation is provided.

Both [123] and [36] create a very similar architecture to approach the integration of IoMT. In [123], the authors develop an integrated medical platform for RHM. The platform is multi-layered to enable easy integration and expandability. The first layer, the perception layer, combines all the sensors that collect data about the patient and their surrounding environment that could affect the patient's health. The next layer is the network and gateway layer, which transfers the data from the perception layer and processes it. The last layer stores and integrates the data received from various sensors and makes informed decisions. The authors showcased the architecture using sensors to measure multiple vitals, including heart rate and body temperature and environmental data like light intensity, humidity, and temperature. They also utilized the system for fall detection. The system is promising but does not present a means of interacting with users. It also presents great resemblance to the architecture used in [36], which augmented it with the mHealth dataset and a UI for caregivers (in the application layer), yet not its primary users, patients.

Most architectures in literature highly depend on cloud infrastructure for storage and analytics. Other architectures include Hadoop Map Reduce techniques to process vast amounts of data in parallel. Although utilizing the cloud provides virtually unlimited space and computational power, a side effect is the time delay in transferring data. This led researchers to utilize faster communication, analysis, and temporary storage within the user's local network to promote a more effective response to dangerous situations with minimum to zero delays.

The devices that provide these functionalities are referred to as fog devices. The name comes from the real-life fog to resemble its proximity to the ground (environment) instead of the cloud. One of the first research works in that area was presented by Vora et al. [118]. The system implemented is used for the detection of activities of daily living based on a radar sensor. The authors highlight further advantages in using fog devices for IoHT. It increases the bandwidth and lowers the latency of data transfer and offers a complementary decrease in energy consumption and data overload compared to cloud computing.

Loza-Matovelle et al. [44] propose an architecture that focuses on integrating robotics and HRI methods. The system combines a network of heterogeneous sensors and actuators in a decentralized manner that decouples the functionalities of various agents. It is made of two servers that communicate together. A local (ROS-based) server oversees task achievement while a server for web services integrates with interactions with the users. All communications in the system use the MQTT protocol. The system is implemented with different sensors, a robotic platform, and a hologram for user interaction, thus showcasing its functionality. While the system shows great promise of integrating various kinds of agents and modularity, it will need a ROS network integration for each new agent, which is not useful for non-robotic agents. Therefore, further breakdown of the local server can prove more practical, as shown in other works discussed here.

Nasr [42] provides a framework that focuses on modularity and scalability and integrates heterogeneous agents, sensors, robots, and HRI devices for AAL scenarios. In [124], the framework is deployed in two scenarios (RHM and AAL), using the same building blocks and combining both a cloud and a fog layer. The framework is divided into three independent layers, namely:

- Device Layer: Includes smart sensors in the environment or on the user's body, robotic agents, and agents used for user interaction such as mobile phones, virtual assistants, or gesture control devices. This layer covers all kinds of objects that can interact with the real world and users.
- IoT Fog Layer: Decentralized computing and storage device(s) that receive the data from the device layer, preprocess it, and sends required data to the cloud. Fog devices reside in the local network of their users.
- Cloud Layer: Central hub for large data storage and analytics. It allows easily expanding the capacity of the system for processing and prediction models.

The authors argue that the proposed division of layers and clearly defined communication protocols and methods allow the framework to be flexible enough to fit into different use cases and personalize each user according to their needs. Two prototypes are created using the framework with different agents and capabilities, but the prototypes still need user testing and feedback in real-world scenarios.

Similarly, a multi-layer architecture made of a device layer, fog computing layer, and a cloud layer is presented [46]. The device and fog layer house the physical devices in the environment and the fog devices, respectively. The cloud layer separates the data and devices from the application and manages the data, contains the rule and data analytics engine. The architecture is implemented in an AAL environment with real-time monitoring of HR and an indoor positioning system. The communication was done through the MQTT protocol. The data is stored using a Redis database in the fog layer and a MongoDB in the cloud.

Feria et al. [45] created an architecture of 3 separate layers. The remote portable device layer combines all the sensing and action devices in the physical world and allows data collection and manipulation of the environment. The devices in this layer contain either a sensor or an actuator, microcontroller, and a communication module, BLE, in their implementation. The second layer receives data from the remote portable device layer and coordinates the devices through adding, removing, or applying changes to devices. The collected data is also manipulated and temporarily saved before being sent to the next layer, and time-sensitive reactions are made at this layer. Therefore, it can be thought of as a fog layer like that proposed in [42], [46], [118]. The final layer is called the Web service application layer. It reorganizes the data to be presented to the users in different forms of user interfaces. The architecture utilizes BLE for communication with devices and JSON and RESTful API for communication with the web service layer. The system is presented to the user as a service-oriented architecture, where the users can interact and control the devices and functionalities in the form of services. The system, however, was not tested in a real-world environment or implemented as a prototype.

## VI. DISCUSSIONS, CHALLENGES, AND FUTURE DIRECTIONS

This section describes the open discussions of the reviewed frameworks and the challenges available in existing systems. In addition, the potential future research directions are highlighted to demonstrate the scope for further study.

### A. OPEN DISCUSSIONS

In this review, we have described smart healthcare frameworks highlighting areas such as health monitoring systems based on wearable devices and smartphones, disease detection using machine learning, utilizing IoMT and social robots for AAL, and software integration architectures used to develop such assistive frameworks.

The summary of the health monitoring systems based on wearable devices and smartphones is illustrated in Tables 1 and 2. From Table 1, it is evident that almost all the systems can measure the heartbeat and body temperature of the patients, which shows the importance of these vitals. Additionally, some of the developed frameworks [54], [55], [56]. Furthermore, [21] measure blood pressure along with the heartbeat and body temperature. The commonly used edge device is NodeMCU in most cases; some of the schemes [51], [52], and [53] used two devices as an edge computing device. Almost all the systems used WiFi for data transfer; only the systems introduced in [55] and [56] utilized Bluetooth for data communication. ThingsPeak and web applications are very common for data visualization that assists the physicians to monitor the patients. Further, Adafruit and LabVIEW are used in [53] and [56] for data visualization. A common negative aspect among the surveyed studies is that the frameworks are not adequately manufactured for clinical uses. From Table 2, it is found that most of the reviewed systems monitored a single sign for the patients except the frameworks developed in [61] and [67]. The camera module (rear and front) is used in most cases as a sensing element; only the schemes introduced in [61] and [64] utilized a microphone and accelerometer for data perception. The different models of iPhone and Samsung brand smartphone are used for the experiment, although the prototype presented in [64] did not mention any smartphone model. The highest number of participants (205 people) are found in [63], and the lowest number (5 individuals) is in [66]. The collected data's maximum and minimum video duration is about 15 min and 10 seconds for [63] and [66], respectively. A data sampling rate of 30 Hz is utilized almost in all cases. No security concerns are handled in most smartphone-based health monitoring systems.

The summary of the machine learning-based disease diagnosis (COVID-19, heart disease, and diabetes) in the IoHT environment is shown in Tables 3, 4, and 5. It is observed from Table 3 that the reviewed systems utilized both the benchmark and real-time sensor data, as the amount of data for COVID-19 cases is relatively small. Most of the systems used CNN or variants of CNN as a classification algorithm; conventional machine learning technique is also used in some cases [81], [28], and [85]. The highest and lowest accuracy of 99.06% and 66.67% are found from [26] and [79]. Almost all the systems are not practically used in the target environment. It is shown from Table 4 that some common benchmark datasets like Hungarian and Cleveland databases, along with real-time sensor data, are used for the experiments. The highest accuracy of 99.45% is achieved from [31] using the MSSO-ANFIS classification technique, and the minimum value of 89% as an accuracy measure is obtained from [94] using bagging ensemble learning. No wearable prototypes are developed in almost all heart disease diagnosis systems. In diabetes detection, only a popular benchmark dataset (Pima Indians Diabetes dataset) is used in [108], and the other frameworks utilized real-time data collected from targeted individuals. Traditional machine learning techniques are applied to develop support systems for diabetes patients. The accuracy values 99.66% and 80.51% are appraised from [105], and [106] are treated as maximum and minimum. The clinical trials of the developed frameworks are absent from the study.

Ambient assisted living has taken a huge portion of researchers' interest in healthcare frameworks and IoT. The review presented here divided the efforts in the field into two portions. First, smart home environments utilize sensors and actuators to assist older adults in living longer, healthier lives while reducing the need for specialized healthcare professionals and its associated cost. Researchers focus on methods to track older adults' motion and identify their activities in their home environments. The systems developed in [34], [114] utilized BLE and RSSI to develop a low-cost localization system and deployed machine learning models for activity recognition [36]. Moreover, unobtrusive sensors have been of increasing importance over recent years. The primary motivation is to present solutions that would require minimal effort from older adults to incorporate into their daily lives. To that end, [37], [115], [117], [118] propose solutions that depend on radar, motion sensors, and door sensors to identify the senior's location, activity and detect dangerous situations such as falls. Finally, approaches that involve more dependency on user interaction, cloud computing, and storage for better predictive analysis are presented in [22], [37], [118], [125].

These studies are all presented as work-alone systems and do not attempt to capture the full-capacity of capabilities IoT presents to this sector. For example, several studies propose utilizing robotic systems to provide social companionship, assistance in activities of daily living, and natural interaction with the provided home systems. Table 6 summarizes the approaches to develop social robots in recent literature. A main component of the provided robots is their ability to autonomously navigate the indoor environment and assist older adults without human interference. Other capabilities requested by various stakeholders include fetch and carry facial recognition and video chatting for telepresence. Researchers developed robots to act as the sole agent for care of the older ad [39]–[41]. Although this brings users personalization, it fails to capture the power of a fully integrated. To that end, [42]–[44], [119] integrate existing commercial robotic agents into IoT environments. The main system behind the robotic systems was ROS, and the main communication protocols utilized were REST and MQTT. The sensors linked to these robotic systems included smartwatches/bracelets and cameras. A fallback of the proposed systems is the lack of real-world validation and extensive testing. Such testing is needed with user feedback to showcase its applicability beyond research and improve the design and functionalities in a user-based approach.

The presented review also highlights the most recent suggestions for software architectures for smart healthcare. The main factor that combines the proposed systems is modularity and multi-layered architecture. The proposed architectures can be summarized to a generic framework as shown in Figure 4, comprised of the main users, a device layer, cloud layer, and user interaction devices. A perception layer or device layer is utilized to combine the sensory data from multiple sources and stream it to the rest of the system for processing. Such a layer is utilized in most recent literature. The collected data is then processed in either a local device or in the cloud. Fog Computing is proposed to accelerate response to real-time data and serve as temporary storage [42], [45], [46], [118]. Furthermore, researchers propose a link to the cloud for greater computational and storage power and enhanced scalability. This leads to the second approach, which combines both a fog layer and a cloud layer, as shown in Figure 5. Several discussed studies have used the fog layer to provide real-time analytics and faster response to dangerous situations. The modularization of smart healthcare is essential, as shown in several studies explored in this paper. However, this emphasizes standardizing communication protocols between the layers, especially between the devices and the device or perception layer. Several studies opted to create their sensors to conform to their proposed architectures, thus decreasing the potential benefit. Conforming to commercial off-the-shelf sensors and reaping the benefits of the wide-spread spectrum of sensors for remote healthcare is crucial, which led studies such as [42], [122] to provide a direct link to incorporate these systems. A major fallback in most software architectures is their abstract nature. Developing market-ready software architectures goes beyond linking the layers and transmitting data. Other aspects such as security and privacy need to be addressed heavily to enable a commercial smart healthcare framework. Such aspects are usually addressed independently from the framework, and the authors see a vital need to include them in the design of the software architecture to enable a complete framework for smart healthcare.

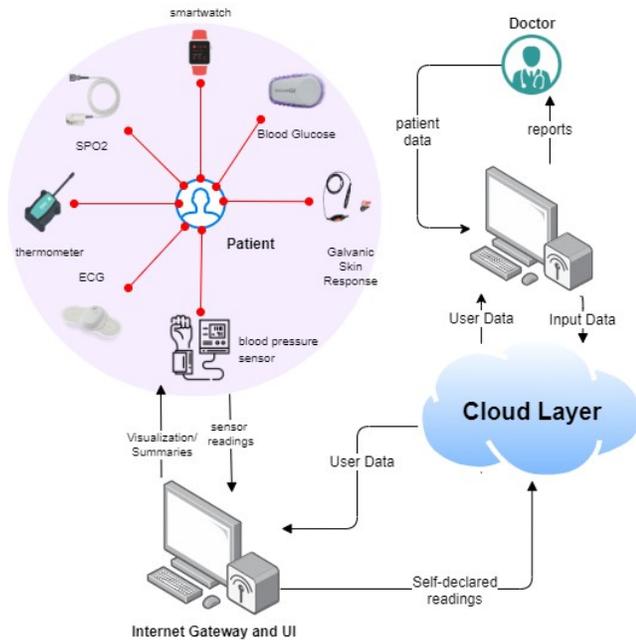

**FIGURE 4.** Generalized Architecture in IoMT - 1.

### B. CHALLENGES AND FURTHER RESEARCH DIRECTIONS

Although several assistive frameworks have been developed using modern technologies to ensure smart healthcare, some challenges need to be addressed to ensure a scalable, secure, easily accessible, and efficient healthcare system. The main challenges, along with the potential future research directions, are demonstrated here.

The major challenge for implementing smart healthcare using wearable devices, including smartphones, is integrating the data from different sensors. As the various sensors generate several data types, it is important to convert the signals from heterogeneous sensors attached to patients to a meaningful format for health monitoring applications. Several data fusion techniques [126]–[128] for integrating information derived from multi-sensory devices can be investigated as a means of providing streamlined signals for improving reliability and minimizing the bandwidth required for communication with the cloud layer as future work. Further, a hybrid body-sensor network architecture based on multi-sensor data fusion approaches will be investigated based on the work of [129]–[131].

Another key issue relating to the healthcare system based on wearable devices is the security and privacy of patients' responsive health records. The security issue has become a widespread and continuous challenge for wearable devices in IoT environments because of the increasing complexity of the data and the progressive network attacks. In the future, more secure and privacy-preserving frameworks using different security ensuring protocols like Blockchain [132]–[134] are recommended that can ensure secure data communication among the users (patients and their families, medical experts, and caregivers). Low power consumption and energy efficiency are very significant for smart healthcare systems based on wearable devices and smartphones, especially for long-term patient monitoring. These issues can be handled by using low power equipment [135], long-life batteries [136], and energy harvesting techniques [137], [138] in future research. Another way to increase the battery's lifetime is the 'sleep' and 'wake up' property of the sensors employed to ensure the desired goal.

Along with the previous challenges, smartphone-based health monitoring systems face some noises as the collected data from smartphone cameras are in image/video format. Generally, the noise in the data delivers the misinformation to the users. Some major developments for healthcare applications are needed that are capable of handling noisy environments [139], [140], or some noise-free solutions [141] as future works. The developed prototypes should be maintained like low-cost, easy-to-use, and compatible platforms to increase the acceptance rate. More research and development activities can be ensured for developing assistive devices for smart healthcare considering the user needs.

Machine learning-based disease diagnosis systems also suffer from various unique challenges that need to be resolved to develop efficient and accurate frameworks for disease detection in IoHT environments.

Remote patient monitoring raises several real-world challenges, such as what to do with missing or incomplete data. Loss of electric power may cause the loss of some data being collected. In the worst case, a natural catastrophe such as an earthquake, weather related event may cause data loss before it is archived at a central cloud location. This would be particularly problematic for patients with serious illness at home. Also challenging is when multiple patients have severe conditions that require assistance beyond what can be responded in a timely way by healthcare teams. There will need to be ways to have a way to send those requests to another healthcare provider. Wearable devices may also fail, and so there may be incomplete and inconsistent data. We will need to have ways to deal with missing data, such as those proposed by Kaur and colleagues [150], [151].

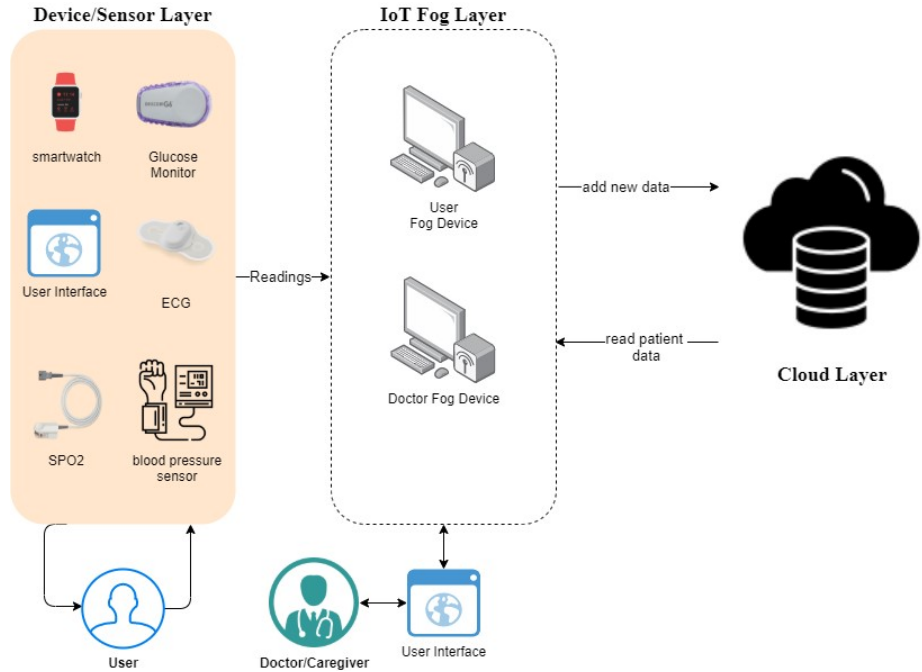

**FIGURE 5.** Generalized Architecture in IoMT - 2.

Another challenge is to fit machine learning and deep learning algorithms with a small amount of data. To resolve the issues related to data shortage, optimized learning algorithms [142], end-to-end architecture [143], and synthetic data generation using Generative adversarial networks [144] are highly recommended as future studies. Furthermore, there are several unnecessary features in the dataset for detecting heart disease responsible for occurring the disease, and these features often degrade the performance of the developed systems. In this scenario, the use of some optimization and feature selection algorithms such as genetic algorithm [145], particle swarm optimization [146], principal component analysis [147], etc., would be a potential solution to improve the performance of the detection procedure in future research. Furthermore, most of the studies depicted that the diabetes detection systems used data from the glucose sensors to achieve their goal. Hence, the proper design of sensors with long life would be an excellent approach for diabetes monitoring and detection. The design of lightweight machine learning frameworks [148], [149] would be better suited in the embedded devices to ensure smart healthcare systems in the future study. Overall, it is found from the reviewed systems that the developed prototypes are not entirely manufactured for practical uses. In some cases, no clinical tests have been conducted yet. Most of the systems represented their results considering the laboratory environment. Addressing these issues considering human health conditions will lead to potential research directions in the future.

AAL is a crucial application for smart healthcare, owing to the benefits it brings in improved and 24-hour monitoring of health and cost reductions on both the older adults and the healthcare system. Nevertheless, there are some challenges left to tackle. First, a user-based approach is required with feedback from older adults and redesign of the system accordingly. The aging society has special needs that are tough to identify without real-life tests and formal user studies. Moreover, the utilization of robotic agents and their integration into a smart environment is important to provide both physical assistance and social presence. Usability and acceptability of smart healthcare systems by older adults is linked to enabling independent living in their own home [36], [43], [109], [111], personalization [36], [37], and the intuitiveness of user-interface as highlighted by [39], [41], [42], [111]. As a result, it is important to directly tackle these three aspects in the design of AAL systems and follow user-based testing and improvement with older adults in the future study.

Software architectures meant for smart healthcare have been improving and following the same direction of modularization and scalability. Recent studies highlighted the need for providing means of integrating COTS and therefore standardizing communication technologies and protocols. Such protocols need to enable multiple user interfaces to accommodate different users and applications. Moreover, both fog devices and cloud integration are needed for real-time response and the power of the cloud for big data analysis, storage, and scalability. Finally, security and privacy concerns are of utmost importance when it comes to health-related data. Consequently, these aspects need to be part of the design of the smart healthcare frameworks at an early stage, utilizing the most recent advancements in Blockchain technologies and allowing access to data appropriate to users in question.

## VII. CONCLUSION

Smart healthcare provides a secure, effective, and easily deployable health monitoring system that can ensure quality healthcare services at a fraction of the cost currently incurred by hospitals or assisted living centers. In this review, we briefly discussed the state-of-the-art wearable devices and smartphones for basic signs monitoring, machine learning for three significant diseases such as (COVID-19, heart disease, and diabetes) diagnosis, and the frameworks developed to aid the adults in ambient assisted living. The software integration frameworks that are very substantial to develop smart healthcare are demonstrated in a nutshell in this review. We have reviewed the advantages and shortcomings of a wide range of systems. In addition, we discussed the major challenges of recently developed smart healthcare frameworks that are the main obstacles to develop assistive prototypes. Some potential future research directions are recommended for the further improvement of the existing healthcare system. It is quite impossible to replace the whole medical system with technology, but it can reduce the burden of medical experts by introducing some novel architectures. The development of such assistive systems would quite feasible while the medical experts and the researchers would work jointly in a platform.

## REFERENCES

[1] World Health Organization, *World report on ageing and health*. World Health Organization, 2015.
[2] D. Calvaresi, D. Cesarini, P. Sernani, M. Marinoni, A. F. Dragoni, and A. Sturm, "Exploring the ambient assisted living domain: A systematic review," *Journal of Ambient Intelligence and Humanized Computing*, vol. 8, no. 2, pp. 239–257, 2017.
[3] A. Charulatha and R. Sujatha, "Smart healthcare use cases and applications," in the *Internet of Things Use Cases for the Healthcare Industry*, Springer, 2020, pp. 185–203.
[4] M. Patton, "Us health care costs rise faster than inflation," *Retrieved from Forbes: https://scholar.google com/scholar*, 2015.
[5] S. B. Baker, W. Xiang, and I. Atkinson, "Internet of things for smart healthcare: Technologies, challenges, and opportunities," *IEEE Access*, vol. 5, pp. 26521–26544, 2017.
[6] A. Rghioui and A. Oumnad, "Challenges and opportunities of internet of things in healthcare," *International Journal of Electrical & Computer Engineering (2088-8708)*, vol. 8, no. 5, 2018.
[7] H. Zhu, C. K. Wu, C. H. Koo, Y. T. Tsang, Y. Liu, H. R. Chi, and K.-F. Tsang, "Smart healthcare in the era of internet-of-things," *IEEE Consumer Electronics Magazine*, vol. 8, no. 5, pp. 26–30, 2019.
[8] S. Dash, S. K. Shakyawar, M. Sharma, and S. Kaushik, "Big data in healthcare: Management, analysis and future prospects," *Journal of Big Data*, vol. 6, no. 1, pp. 1–25, 2019.
[9] S. Bahri, N. Zoghlami, M. Abed, and J. M. R. Tavares, "Big data for healthcare: A survey," *IEEE Access*, vol. 7, pp. 7397–7408, 2018.
[10] O. Ali, A. Shrestha, J. Soar, and S. F. Wamba, "Cloud computing-enabled healthcare opportunities, issues, and applications: A systematic review," *International Journal of Information Management*, vol. 43, pp. 146–158, 2018.
[11] F. Gao and A. Sunyaev, "Context matters: A review of the determinant factors in the decision to adopt cloud computing in healthcare," *International Journal of Information Management*, vol. 48, pp. 120– 138, 2019.
[12] E. Wood, D. Mohamedally, N. J. Sebire, and S. Visram, *44 internet of healthcare things (ioht) handheld device for secure patient data retrieval*, 2020.
[13] H. Kaur, M. Atif, and R. Chauhan, "An internet of healthcare things (ioht)-based healthcare monitoring system," in *Advances in intelligent computing and communication*, Springer, 2020, pp. 475–482.
[14] G. J. Joyia, R. M. Liaqat, A. Farooq, and S. Rehman, "Internet of medical things (iomt): Applications, benefits and future challenges in healthcare domain," *J Commun*, vol. 12, no. 4, pp. 240–247, 2017.
[15] V. S. Parvathy, S. Pothiraj, and J. Sampson, "Automated internet of medical things (iomt) based healthcare monitoring system," in *Cognitive Internet of Medical Things for Smart Healthcare*, Springer, 2021, pp. 117–128.
[16] S. Majumder, T. Mondal, and M. J. Deen, "Wearable sensors for remote health monitoring," *Sensors*, vol. 17, no. 1, p. 130, 2017.
[17] D. Dias and J. Paulo Silva Cunha, "Wearable health devices—vital sign monitoring, systems and technologies," *Sensors*, vol. 18, no. 8, p. 2414, 2018.
[18] M. M. Islam, S. Mahmud, L. J. Muhammad, M. R. Islam, S. Nooruddin, and S. I. Ayon, "Wearable technology to assist the patients infected with novel coronavirus (covid-19)," *SN Computer Science*, vol. 1, no. 6, pp. 1–9, 2020. DOI: 10.1007/s42979-02000335-4.
[19] F. J. Dian, R. Vahidnia, and A. Rahmati, "Wearables and the internet of things (iot), applications, opportunities, and challenges: A survey," *IEEE Access*, vol. 8, pp. 69200–69211, 2020.
[20] M. M. Islam, A. Rahaman, and M. R. Islam, "Development of smart healthcare monitoring system in iot environment," *SN Computer Science*, vol. 1, no. 3, pp. 1–11, 2020. DOI: 10.1007/s42979-020-00195-y.
[21] J. Wan, M. A. Al-awlaqi, M. Li, M. O'Grady, X. Gu, J. Wang, and N. Cao, "Wearable iot enabled real-time health monitoring system," *EURASIP Journal on Wireless Communications and Networking*, vol. 2018, no. 1, p. 298, 2018.
[22] Y. Quintana, F. Darren, B. Crotty, J. Ruchira, E. Kaldany, M. Gorenberg, L. Lipsitz, D. Engorn, J. Rodriguez, A. Orfanos, *et al.*, "Infosage: Supporting elders and families through online family networks," in *AMIA Annual Symposium Proceedings*, American Medical Informatics Association, vol. 2018, 2018, p. 932.
[23] S. Majumder and M. J. Deen, "Smartphone sensors for health monitoring and diagnosis," *Sensors*, vol. 19, no. 9, p. 2164, 2019.
[24] G. Zhang, Z. Mei, Y. Zhang, X. Ma, B. Lo, D. Chen, and Y. Zhang, "A noninvasive blood glucose monitoring system based on smartphone ppg signal processing and machine learning," *IEEE Transactions on Industrial Informatics*, vol. 16, no. 11, pp. 7209–7218, 2020.
[25] F. Tabei, J. M. Gresham, B. Askarian, K. Jung, and J. W. Chong, "Cuff-less blood pressure monitoring system using smartphones," *IEEE Access*, vol. 8, pp. 11534–11545, 2020.
[26] D.-N. Le, V. S. Parvathy, D. Gupta, A. Khanna, J. J. Rodrigues, and K. Shankar, "Iot enabled depthwise separable convolution neural network with deep support vector machine for covid-19 diagnosis and classification," *International Journal of Machine Learning and Cybernetics*, vol. 12, no. 1, pp. 1–14, 2021. DOI: 10.1007/s13042-020-01248-7.
[27] S. Karmore, R. Bodhe, F. Al-Turjman, R. L. Kumar, and S. Pillai, "Iot based humanoid software for identification and diagnosis of covid-19 suspects," *IEEE Sensors Journal*, pp. 1–1, 2020. DOI: 10.1109/JSEN. 2020.3030905.
[28] D. Cacovean, I. Ioana, and G. Nitulescu, "Iot system in diagnosis of covid-19 patients," *Informatica Economica*, vol. 24, no. 2, pp. 75–89, 2020.
[29] F. Ali, S. El-Sappagh, S. R. Islam, D. Kwak, A. Ali, M. Imran, and K.-S. Kwak, "A smart healthcare monitoring system for heart

disease prediction based on ensemble deep learning and feature fusion," *Information Fusion*, vol. 63, pp. 208–222, 2020.

[30] O. Deperlioglu, U. Kose, D. Gupta, A. Khanna, and A. K. Sangaiah, "Diagnosis of heart diseases by a secure internet of health things system based on autoencoder deep neural network," *Computer Communications*, vol. 162, pp. 31–50, 2020.

[31] M. A. Khan and F. Algarni, "A healthcare monitoring system for the diagnosis of heart disease in the iomt cloud environment using msso-anfis," *IEEE Access*, vol. 8, pp. 122259–122269, 2020.

[32] A. Rghioui, A. Naja, J. L. Mauri, and A. Oumnad, "An iot based diabetic patient monitoring system using machine learning and node mcu," in *Journal of Physics: Conference Series*, IOP Publishing, vol. 1743, 2021, p. 012035.

[33] P. P. Ponmalar and C. Vijayalakshmi, "Aggregation in iot for prediction of diabetics with machine learning techniques," in *International conference on Computer Networks, Big data and IoT*, Springer, 2019, pp. 789–798.

[34] K. Komai, M. Fujimoto, Y. Arakawa, H. Suwa, Y. Kashimoto, and K. Yasumoto, "Beacon-based multiperson activity monitoring system for day care center," in *2016 IEEE International Conference on Pervasive Computing and Communication Workshops (PerCom Workshops)*, 2016, pp. 1–6. DOI: 10.1109/PERCOMW.2016.7457140.

[35] G. Marques, I. M. Pires, N. Miranda, and R. Pitarma, "Air quality monitoring using assistive robots for ambient assisted living and enhanced living environments through internet of things," *Electronics*, vol. 8, no. 12, p. 1375, 2019.

[36] L. Syed, S. Jabeen, S. Manimala, and A. Alsaeedi, "Smart healthcare framework for ambient assisted living using iomt and big data analytics techniques," *Future Generation Computer Systems*, vol. 101, pp. 136–151, 2019.

[37] S. Stavrotheodoros, N. Kaklanis, and D. Tzovaras, "A personalized cloud-based platform for aal support to cognitively impaired elderly people," in *International Conference on Biomedical and Health Informatics*, Springer, 2017, pp. 87–91.

[38] S. Bedaf, P. Marti, F. Amirabdollahian, and L. de Witte, "A multi-perspective evaluation of a service robot for seniors: The voice of different stakeholders," *Disability and Rehabilitation: Assistive Technology*, vol. 13, no. 6, pp. 592–599, 2018.

[39] S. Lee and A. M. Naguib, "Toward a sociable and dependable elderly care robot: Design, implementation and user study," *Journal of Intelligent & Robotic Systems*, vol. 98, no. 1, pp. 5–17, 2020.

[40] D. Portugal, L. Santos, P. Alvito, J. Dias, G. Samaras, and E. Christodoulou, "Socialrobot: An interactive mobile robot for elderly home care," in *2015 IEEE/SICE International Symposium on System Integration (SII)*, IEEE, 2015, pp. 811–816.

[41] H.-M. Gross, A. Scheidig, S. Müller, B. Schütz, C. Fricke, and S. Meyer, "Living with a mobile companion robot in your own apartment-final implementation and results of a 20-weeks field study with 20 seniors," in *2019 International Conference on Robotics and Automation (ICRA)*, IEEE, 2019, pp. 2253–2259.

[42] M. Nasr, F. Karray, and Y. Quintana, "Human machine interaction platform for home care support system," in *2020 IEEE International Conference on Systems, Man, and Cybernetics (SMC)*, IEEE, 2020, pp. 4210–4215.

[43] H. M. Do, M. Pham, W. Sheng, D. Yang, and M. Liu, "Rish: A robot-integrated smart home for elderly care," *Robotics and Autonomous Systems*, vol. 101, pp. 74–92, 2018.

[44] D. Loza-Matovelle, A. Verdugo, E. Zalama, and J. Gómez-Garcıa-Bermejo, "An architecture for the integration of robots and sensors for the care of the elderly in an ambient assisted living environment," *Robotics*, vol. 8, no. 3, p. 76, 2019.

[45] F. Feria, O. J. S. Parra, and B. S. R. Daza, "Design of an architecture for medical applications in iot," in *International Conference on Cooperative Design, Visualization and Engineering*, Springer, 2016, pp. 263–270.

[46] B. Farahani, F. Firouzi, V. Chang, M. Badaroglu, N. Constant, and K. Mankodiya, "Towards fogdriven iot ehealth: Promises and challenges of iot in medicine and healthcare," *Future Generation Computer Systems*, vol. 78, pp. 659–676, 2018.

[47] S. Gahlot, S. Reddy, and D. Kumar, "Review of smart health monitoring approaches with survey analysis and proposed framework," *IEEE Internet of Things Journal*, vol. 6, no. 2, pp. 2116–2127, 2018.

[48] M. Talal, A. Zaidan, B. Zaidan, A. Albahri, A. Alamoodi, O. Albahri, M. Alsalem, C. Lim, K. L. Tan, W. Shir, *et al.*, "Smart home-based iot for real-time and secure remote health monitoring of triage and priority system using body sensors: Multi-driven systematic review," *Journal of medical systems*, vol. 43, no. 3, p. 42, 2019.

[49] A. Rahaman, M. M. Islam, M. R. Islam, M. S. Sadi, and S. Nooruddin, "Developing iot based smart health monitoring systems: A review," *Revue d'Intelligence Artificielle*, vol. 33, no. 6, pp. 435– 440, 2019.

[50] K. Elango and K. Muniandi, "A low-cost wearable remote healthcare monitoring system," *Role of Edge Analytics in Sustainable Smart City Development: Challenges and Solutions*, pp. 219–242, 2020.

[51] M. A. Al-Sheikh and I. A. Ameen, "Design of mobile healthcare monitoring system using iot technology and cloud computing," in *IOP Conference Series: Materials Science and Engineering*, IOP Publishing, vol. 881, 2020, p. 012113.

[52] A. Chigozirim, N. O. Vivian, N. J. Uchenna, and A. A. Oreoluwa, "A patient monitoring system using internet of things technology," *Ingénierie des Systèmes d'Information*, vol. 25, no. 3, pp. 351–357, 2020.

[53] S. Mohapatra, S. Mohanty, and S. Mohanty, "Smart healthcare: An approach for ubiquitous healthcare management using iot," in *Big Data Analytics for Intelligent Healthcare Management*, Elsevier, 2019, pp. 175–196.

[54] K. N. Swaroop, K. Chandu, R. Gorrepotu, and S. Deb, "A health monitoring system for vital signs using iot," *Internet of Things*, vol. 5, pp. 116–129, 2019.

[55] M. Al-Khafajiy, T. Baker, C. Chalmers, M. Asim, H. Kolivand, M. Fahim, and A. Waraich, "Remote health monitoring of elderly through wearable sensors," *Multimedia Tools and Applications*, vol. 78, no. 17, pp. 24681–24706, 2019.

[56] N. Semwal, M. Mukherjee, C. Raj, and W. Arif, "An iot based smart e-health care system," *Journal of Information and Optimization Sciences*, vol. 40, no. 8, pp. 1787–1800, 2019.

[57] A. Kumar, G. Chattree, and S. Periyasamy, "Smart healthcare monitoring system," *Wireless Personal Communications*, vol. 101, no. 1, pp. 453–463, 2018.

[58] R. Edirisinghe, A. Stranieri, and N. Wickramasinghe, "A taxonomy for mhealth," in *Virtual and Mobile Healthcare: Breakthroughs in Research and Practice*, IGI Global, 2020, pp. 823–842.

[59] A. Bassi, O. John, D. Praveen, P. K. Maulik, R. Panda, and V. Jha, "Current status and future directions of mhealth interventions for health system strengthening in india: Systematic review," *JMIR mHealth and uHealth*, vol. 6, no. 10, e11440, 2018.

[60] D. Albert, B. R. Satchwell, and K. N. Barnett, *Heart monitoring system usable with a smartphone or computer*, US Patent 9,649,042, May 2017.

[61] A. Nemcova, I. Jordanova, M. Varecka, R. Smisek, L. Marsanova, L. Smital, and M. Vitek, "Monitoring of heart rate, blood oxygen saturation, and blood pressure using a smartphone," *Biomedical Signal Processing and Control*, vol. 59, p. 101928, 2020.

[62] M. Alafeef and M. Fraiwan, "Smartphone-based respiratory rate estimation using photoplethysmographic imaging and discrete wavelet transform," *Journal of Ambient Intelligence and Humanized Computing*, vol. 11, no. 2, pp. 693–703, 2020.

[63] J. Dey, A. Gaurav, and V. N. Tiwari, "Instabp: Cuffless blood pressure monitoring on smartphone using single ppg sensor," in

*2018 40th Annual International Conference of the IEEE Engineering in Medicine and Biology Society (EMBC)*, IEEE, 2018, pp. 5002–5005.

[64] E. J. Wang, J. Zhu, M. Jain, T.-J. Lee, E. Saba, L. Nachman, and S. N. Patel, "Seismo: Blood pressure monitoring using built-in smartphone accelerometer and camera," in *Proceedings of the 2018 CHI Conference on Human Factors in Computing Systems*, 2018, pp. 1–9. DOI: 10.1145/3173574.3173999.

[65] M. Alafeef, "Smartphone-based photoplethysmographic imaging for heart rate monitoring," *Journal of Medical Engineering & Technology*, vol. 41, no. 5, pp. 387–395, 2017.

[66] J.-P. Lomaliza and H. Park, "A highly efficient and reliable heart rate monitoring system using smart-phone cameras," *Multimedia Tools and Applications*, vol. 76, no. 20, pp. 21051–21071, 2017.

[67] A. Qayyum, A. S. Malik, A. N. Shuaibu, and N. Nasir, "Estimation of non-contact smartphone videobased vital sign monitoring using filtering and standard color conversion techniques," in *2017 IEEE Life Sciences Conference (LSC)*, IEEE, 2017, pp. 202–205.

[68] M. A. Santos, R. Munoz, R. Olivares, P. P. Rebouças Filho, J. Del Ser, and V. H. C. de Albuquerque, "Online heart monitoring systems on the internet of health things environments: A survey, a reference model and an outlook," *Information Fusion*, vol. 53, pp. 222–239, 2020.

[69] C. A. da Costa, C. F. Pasluosta, B. Eskofier, D. B. da Silva, and R. da Rosa Righi, "Internet of health things: Toward intelligent vital signs monitoring in hospital wards," *Artificial intelligence in medicine*, vol. 89, pp. 61–69, 2018.

[70] J. Manyika, M. Chui, P. Bisson, J. Woetzel, R. Dobbs, J. Bughin, and D. Aharon. Unlocking the potential of the internet of things, [Online]. Available: https://www.mckinsey.com/business-functions/mckinseydigital/our-insights/the-internet-of-things-thevalue-of-digitizing-the-physical-world. Accessed: Jan. 16, 2021.

[71] W. Li, Y. Chai, F. Khan, S. R. U. Jan, S. Verma, V. G. Menon, X. Li, *et al.*, "A comprehensive survey on machine learning-based big data analytics for iotenabled smart healthcare system," *Mobile Networks and Applications*, pp. 1–19, 2021. DOI: 10.1007/s11036-020-01700-6.

[72] F. Samie, L. Bauer, and J. Henkel, "From cloud down to things: An overview of machine learning in internet of things," *IEEE Internet of Things Journal*, vol. 6, no. 3, pp. 4921–4934, 2019.

[73] K. Pradhan and P. Chawla, "Medical internet of things using machine learning algorithms for lung cancer detection," *Journal of Management Analytics*, vol. 7, no. 4, pp. 591–623, 2020.

[74] Worldometer. Covid-19 coronavirus pandemic, [Online]. Available: https://www.worldometers.info/coronavirus/. Accessed: Jan. 22, 2021.

[75] X. Mei, H.-C. Lee, K.-y. Diao, M. Huang, B. Lin, C. Liu, Z. Xie, Y. Ma, P. M. Robson, M. Chung, *et al.*, "Artificial intelligence–enabled rapid diagnosis of patients with covid-19," *Nature medicine*, vol. 26, no. 8, pp. 1224–1228, 2020.

[76] M. M. Islam, F. Karray, R. Alhajj, and J. Zeng, "A review on deep learning techniques for the diagnosis of novel coronavirus (covid-19)," *IEEE Access*, pp. 1–1, 2021. DOI: 10.1109/ACCESS.2021.3058537.

[77] A. Asraf, M. Z. Islam, M. R. Haque, and M. M. Islam, "Deep learning applications to combat novel coronavirus (covid-19) pandemic," *SN Computer Science*, vol. 1, no. 6, pp. 1–7, 2020. DOI: 10.1007/s42979-020-00383-w.

[78] J. P. Cohen, P. Morrison, L. Dao, K. Roth, T. Q. Duong, and M. Ghassemi, "Covid-19 image data collection: Prospective predictions are the future," *arXiv 2006.11988*, 2020. [Online]. Available: https://github.com/ieee8023/covid-chestxray-dataset.

[79] A. P. Ramallo-González, A. González-Vidal, and A. F. Skarmeta, "Ciotvid: Towards an open iotplatform for infective pandemic diseases such as covid-19," *Sensors*, vol. 21, no. 2, p. 484, 2021.

[80] I. Ahmed, A. Ahmad, and G. Jeon, "An iot based deep learning framework for early assessment of covid-19," *IEEE Internet of Things Journal*, pp. 1–1, 2020. DOI: 10.1109/JIOT.2020.3034074.

[81] M. Otoom, N. Otoum, M. A. Alzubaidi, Y. Etoom, and R. Banihani, "An iot-based framework for early identification and monitoring of covid-19 cases," *Biomedical Signal Processing and Control*, vol. 62, p. 102149, 2020.

[82] Covid-19 open research dataset (cord-19). 2020. version 2020-03-13, [Online]. Available: https://pages.semanticscholar.org/coronavirus-research. Accessed: Mar. 22, 2020.

[83] N. El-Rashidy, S. El-Sappagh, S. Islam, H. M. ElBakry, and S. Abdelrazek, "End-to-end deep learning framework for coronavirus (covid-19) detection and monitoring," *Electronics*, vol. 9, no. 9, p. 1439, 2020.

[84] A. Altun and O. Erdogan, "A chest x-ray image: Total opacification of the hemithorax.," *Cardiology in review*, vol. 11, no. 6, pp. 301–302, 2003.

[85] K. Kumar, M. Iyapparaja, V. Niveditha, S. Magesh, G. Magesh, and S. Marappan, "Monitoring and analysis of the recovery rate of covid-19 positive cases to prevent dangerous stage using iot and sensors," *International Journal of Pervasive Computing and Communications*, 2020. DOI: 10.1108/IJPCC-072020-0088.

[86] H. Ahmed, E. M. Younis, A. Hendawi, and A. A. Ali, "Heart disease identification from patients' social posts, machine learning solution on spark," *Future Generation Computer Systems*, vol. 111, pp. 714–722, 2020.

[87] S. Kaptoge, L. Pennells, D. De Bacquer, M. T. Cooney, M. Kavousi, G. Stevens, L. M. Riley, S. Savin, T. Khan, S. Altay, *et al.*, "World health organization cardiovascular disease risk charts: Revised models to estimate risk in 21 global regions," *The Lancet Global Health*, vol. 7, no. 10, e1332–e1345, 2019.

[88] S. S. Sarmah, "An efficient iot-based patient monitoring and heart disease prediction system using deep learning modified neural network," *IEEE Access*, vol. 8, pp. 135784–135797, 2020.

[89] Heart disease data set, [Online]. Available: https://archive.ics.uci.edu/ml/datasets/heart+Disease. Accessed: Jan. 15, 2020.

[90] W. Zhang, J. Han, and S. Deng, "Heart sound classification based on scaled spectrogram and tensor decomposition," *Expert Systems with Applications*, vol. 84, pp. 220–231, 2017.

[91] H. Coskun, T. Yigit, and O. Deperlioglu, "Effect of filter selection on classification of extrasystole heart sounds via mobile devices," in *2016 IEEE 10th International Conference on Application of Information and Communication Technologies (AICT)*, IEEE, 2016, pp. 1–5.

[92] Framingham heart study dataset, [Online]. Available: https://www.kaggle.com/amanajmera1/framingham-heart-study-dataset. Accessed: Jan. 15, 2020.

[93] M. A. Khan, "An iot framework for heart disease prediction based on mdcnn classifier," *IEEE Access*, vol. 8, pp. 34717–34727, 2020.

[94] S. Tuli, N. Basumatary, S. S. Gill, M. Kahani, R. C. Arya, G. S. Wander, and R. Buyya, "Healthfog: An ensemble deep learning based smart healthcare system for automatic diagnosis of heart diseases in integrated iot and fog computing environments," *Future Generation Computer Systems*, vol. 104, pp. 187–200, 2020.

[95] T.-H. Nguyen, T.-N. Nguyen, and T.-T. Nguyen, "A deep learning framework for heart disease classification in an iots-based system," in *A Handbook of Internet of Things in Biomedical and Cyber Physical System*, Springer, 2020, pp. 217–244.

[96] M. Ganesan and N. Sivakumar, "Iot based heart disease prediction and diagnosis model for healthcare using machine learning models," in *2019 IEEE International Conference on System, Computation, Automation and Networking (ICSCAN)*, 2019, pp. 1–5. DOI: 10.1109/ICSCAN.2019.8878850.

[97] Statlog (heart) data set, [Online]. Available: http://archive.ics.uci.edu/ml/datasets/statlog+ (heart). Accessed: Jan. 20, 2019.

[98] P. Saeedi, I. Petersohn, P. Salpea, B. Malanda, S. Karuranga, N. Unwin, S. Colagiuri, L. Guariguata, A. A. Motala, K. Ogurtsova, *et al.*, "Global and regional diabetes prevalence estimates for 2019 and projections for 2030 and 2045: Results from the international diabetes federation diabetes atlas," *Diabetes research and clinical practice*, vol. 157, p. 107843, 2019.

[99] I. Rodriguez, "On the management of type 1 diabetes mellitus with iot devices and ml techniques," *arXiv: 2101.02409*, 2021. [Online]. Available: https://arxiv.org/abs/2101.02409.

[100] P. Bide and A. Padalkar, "Survey on diabetes mellitus and incorporation of big data, machine learning and iot to mitigate it," in *2020 6th International Conference on Advanced Computing and Communication Systems (ICACCS)*, 2020, pp. 1–10. DOI: 10.1109/ICACCS48705.2020.9074202.

[101] R. Biswas, S. Pal, N. H. H. Cuong, and A. Chakrabarty, "A novel iot-based approach towards diabetes prediction using big data," in *Intelligent Computing in Engineering*, Springer, 2020, pp. 163–170.

[102] V. R. Allugunti, C. K. K. Reddy, N. Elango, and P. Anisha, "Prediction of diabetes using internet of things (iot) and decision trees: Sldps," in *Intelligent Data Engineering and Analytics*, Springer, 2020, pp. 453–461.

[103] M. I. A. Efat, S. Rahman, and T. Rahman, "Iot based smart health monitoring system for diabetes patients using neural network," in *International Conference on Cyber Security and Computer Science*, Springer, 2020, pp. 593–606.

[104] A. Rghioui, A. Naja, and A. Oumnad, "Diabetic patients monitoring and data classification using iot application," in *2020 International Conference on Electrical and Information Technologies (ICEIT)*, 2020, pp. 1–6. DOI: 10.1109/ICEIT48248.2020.9113171.

[105] A. Rghioui, J. Lloret, S. Sendra, and A. Oumnad, "A smart architecture for diabetic patient monitoring using machine learning algorithms," in *Healthcare*, Multidisciplinary Digital Publishing Institute, vol. 8, 2020, p. 348.

[106] B. Godi, S. Viswanadham, A. S. Muttipati, O. P. Samantray, and S. R. Gadiraju, "E-healthcare monitoring system using iot with machine learning approaches," in *2020 International Conference on Computer Science, Engineering and Applications (ICCSEA)*, 2020, pp. 1–5. DOI: 10.1109/ICCSEA49143.2020.9132937.

[107] Pima indians diabetes database, [Online]. Available: https://www.kaggle.com/uciml/pima-indiansdiabetes-database. Accessed: May. 15, 2020.

[108] P. Kaur, N. Sharma, A. Singh, and B. Gill, "Ci-dpf: A cloud iot based framework for diabetes prediction," in *2018 IEEE 9th Annual Information Technology, Electronics and Mobile Communication Conference (IEMCON)*, IEEE, 2018, pp. 654–660.

[109] G. Marques and R. Pitarma, "An indoor monitoring system for ambient assisted living based on internet of things architecture," *International journal of environmental research and public health*, vol. 13, no. 11, p. 1152, 2016.

[110] P. Simoens, M. Dragone, and A. Saffiotti, "The internet of robotic things: A review of concept, added value and applications," *International Journal of Advanced Robotic Systems*, vol. 15, no. 1, p. 1729881418759424, 2018.

[111] D. Portugal, P. Alvito, E. Christodoulou, G. Samaras, and J. Dias, "A study on the deployment of a service robot in an elderly care center," *International Journal of Social Robotics*, vol. 11, no. 2, pp. 317–341, 2019.

[112] S. Yusif, J. Soar, and A. Hafeez-Baig, "Older people, assistive technologies, and the barriers to adoption: A systematic review," *International journal of medical informatics*, vol. 94, pp. 112–116, 2016.

[113] G. Toms, F. Verity, and A. Orrell, "Social care technologies for older people: Evidence for instigating a broader and more inclusive dialogue," *Technology in Society*, vol. 58, p. 101111, 2019.

[114] T. Tegou, I. Kalamaras, M. Tsipouras, N. Giannakeas, K. Votis, and D. Tzovaras, "A low-cost indoor activity monitoring system for detecting frailty in older adults," *Sensors*, vol. 19, no. 3, p. 452, 2019.

[115] G. Diraco, A. Leone, and P. Siciliano, "A radarbased smart sensor for unobtrusive elderly monitoring in ambient assisted living applications," *Biosensors*, vol. 7, no. 4, p. 55, 2017.

[116] O. Banos, R. Garcia, J. A. Holgado-Terriza, M. Damas, H. Pomares, I. Rojas, A. Saez, and C. Villalonga, "Mhealthdroid: A novel framework for agile development of mobile health applications," in *International workshop on ambient assisted living*, Springer, 2014, pp. 91–98.

[117] S. Stavrotheodoros, N. Kaklanis, K. Votis, and D. Tzovaras, "A smart-home iot infrastructure for the support of independent living of older adults," in *IFIP international conference on artificial intelligence applications and innovations*, Springer, 2018, pp. 238–249.

[118] J. Vora, S. Tanwar, S. Tyagi, N. Kumar, and J. J. P. C. Rodrigues, "Faal: Fog computing-based patient monitoring system for ambient assisted living," in *2017 IEEE 19th International Conference on e-Health Networking, Applications and Services (Healthcom)*, 2017, pp. 1–6. DOI: 10.1109/HealthCom.2017.8210825.

[119] F. Gomez-Donoso, F. Escalona, F. M. Rivas, J. M. Cañas, and M. Cazorla, "Enhancing the ambient assisted living capabilities with a mobile robot," *Computational Intelligence and Neuroscience*, vol. 2019, p. 9412384, 2019.

[120] J. Quintas, G. S. Martins, L. Santos, P. Menezes, and J. Dias, "Toward a context-aware human–robot interaction framework based on cognitive development," *IEEE Transactions on Systems, Man, and Cybernetics: Systems*, vol. 49, no. 1, pp. 227–237, 2018.

[121] A. Vercelli, I. Rainero, L. Ciferri, M. Boido, and F. Pirri, "Robots in elderly care," *DigitCult-Scientific Journal on Digital Cultures*, vol. 2, no. 2, pp. 37–50, 2018.

[122] G. Petrovic, V. Dimitrieski, and H. Fujita, "Cloud-´ based health monitoring system based on commercial off-the-shelf hardware," in *2016 IEEE International Conference on Systems, Man, and Cybernetics (SMC)*, IEEE, 2016, pp. 003713–003718.

[123] A. Rashed, A. Ibrahim, A. Adel, B. Mourad, A. Hatem, M. Magdy, N. Elgaml, and A. Khattab, "Integrated iot medical platform for remote healthcare and assisted living," in *2017 Japan-Africa Conference on Electronics, Communications and Computers (JACECC)*, IEEE, 2017, pp. 160–163.

[124] M. Nasr, "Scalable human-machine interaction system for real-time care in the internet of health things," Master's thesis, University of Waterloo, 2020.

[125] E. K. Jaina, L. Lipsitz, D. Engorn, J. Rodriguez, F. Pandolfe, A. Bajracharya, W. v. Slack, and C. Sarfan, "Infosage: Use of online technologies for communication and elder care," *Building Capacity for Health Informatics in the Future*, vol. 234, p. 280, 2017.

[126] R.-T. Wu and M. R. Jahanshahi, "Data fusion approaches for structural health monitoring and system identification: Past, present, and future," *Structural Health Monitoring*, vol. 19, no. 2, pp. 552–586, 2020.

[127] A. P. Kene and S. K. Choudhury, "Analytical modeling of tool health monitoring system using multiple sensor data fusion approach in hard machining," *Measurement*, vol. 145, pp. 118–129, 2019.

[128] R. C. King, E. Villeneuve, R. J. White, R. S. Sherratt, W. Holderbaum, and W. S. Harwin, "Application of data fusion techniques and technologies for wearable health monitoring," *Medical engineering & physics*, vol. 42, pp. 1–12, 2017. DOI: 10.1016/j.medengphy.2016.12.011.

[129] K. Lin, Y. Li, J. Sun, D. Zhou, and Q. Zhang, "Multisensor fusion for body sensor network in medical human–robot interaction scenario," *Information Fusion*, vol. 57, pp. 15–26, 2020.

[130] M. Wang, X. Wang, L. T. Yang, X. Deng, and L. Yi, "Multi-sensor fusion based intelligent sensor relocation for health and safety monitoring in bsns," *Information Fusion*, vol. 54, pp. 61–71, 2020.

[131] R. Dautov, S. Distefano, and R. Buyya, "Hierarchical data fusion for smart healthcare," *Journal of Big Data*, vol. 6, no. 1, pp. 1–23, 2019. DOI: 10.1186/ s40537-019-0183-6.

[132] J. J. Hathaliya and S. Tanwar, "An exhaustive survey on security and privacy issues in healthcare 4.0," *Computer Communications*, vol. 153, pp. 311–335, 2020.

[133] A. Saha, R. Amin, S. Kunal, S. Vollala, and S. K. Dwivedi, "Review on "blockchain technology based medical healthcare system with privacy issues"," *Security and Privacy*, vol. 2, no. 5, e83, 2019.

[134] X. Guo, H. Lin, Y. Wu, and M. Peng, "A new data clustering strategy for enhancing mutual privacy in healthcare iot systems," *Future Generation Computer Systems*, vol. 113, pp. 407–417, 2020.

[135] H. Kalantarian, C. Sideris, B. Mortazavi, N. Alshurafa, and M. Sarrafzadeh, "Dynamic computation offloading for low-power wearable health monitoring systems," *IEEE Transactions on Biomedical Engineering*, vol. 64, no. 3, pp. 621–628, 2016.

[136] D. Zhou, Z. Li, J. Zhu, H. Zhang, and L. Hou, "State of health monitoring and remaining useful life prediction of lithium-ion batteries based on temporal convolutional network," *IEEE Access*, vol. 8, pp. 53307–53320, 2020.

[137] M. Babar, A. Rahman, F. Arif, and G. Jeon, "Energyharvesting based on internet of things and big data analytics for smart health monitoring," *Sustainable Computing: Informatics and Systems*, vol. 20, pp. 155–164, 2018.

[138] F. Mansourkiaie, L. S. Ismail, T. M. Elfouly, and M. H. Ahmed, "Maximizing lifetime in wireless sensor network for structural health monitoring with and without energy harvesting," *IEEE Access*, vol. 5, pp. 2383–2395, 2017.

[139] D. Pollreisz and N. TaheriNejad, "Detection and removal of motion artifacts in ppg signals," *Mobile Networks and Applications*, pp. 1–11, 2019. DOI: 10. 1007/s11036-019-01323-6.

[140] S. K. Bashar, D. Han, A. Soni, D. D. McManus, and K. H. Chon, "Developing a novel noise artifact detection algorithm for smartphone ppg signals: Preliminary results," in *2018 IEEE EMBS International Conference on Biomedical & Health Informatics (BHI)*, IEEE, 2018, pp. 79–82.

[141] D. Yang, Y. Cheng, J. Zhu, D. Xue, G. Abt, H. Ye, and Y. Peng, "A novel adaptive spectrum noise cancellation approach for enhancing heartbeat rate monitoring in a wearable device," *IEEE Access*, vol. 6, pp. 8364–8375, 2018.

[142] Y. Oh, S. Park, and J. C. Ye, "Deep learning covid-19 features on cxr using limited training data sets," *IEEE Transactions on Medical Imaging*, vol. 39, no. 8, pp. 2688–2700, 2020.

[143] J. Song, H. Wang, Y. Liu, W. Wu, G. Dai, Z. Wu, P. Zhu, W. Zhang, K. W. Yeom, and K. Deng, "Endto-end automatic differentiation of the coronavirus disease 2019 (covid-19) from viral pneumonia based on chest ct," *European journal of nuclear medicine and molecular imaging*, vol. 47, no. 11, pp. 2516–2524, 2020.

[144] A. Harshavardhan, H. Bhukya, A. K. Prasad, *et al.*, "Advanced machine learning-based analytics on covid-19 data using generative adversarial networks," *Materials Today: Proceedings*, 2020. DOI: 10.1016/ j.matpr.2020.10.053.

[145] G. T. Reddy, M. P. K. Reddy, K. Lakshmanna, D. S. Rajput, R. Kaluri, and G. Srivastava, "Hybrid genetic algorithm and a fuzzy logic classifier for heart disease diagnosis," *Evolutionary Intelligence*, vol. 13, no. 2, pp. 185–196, 2020.

[146] A. H. Alkeshuosh, M. Z. Moghadam, I. Al Mansoori, and M. Abdar, "Using pso algorithm for producing best rules in diagnosis of heart disease," in *2017 international conference on computer and applications (ICCA)*, IEEE, 2017, pp. 306–311.

[147] A. K. Gárate-Escamila, A. H. El Hassani, and E. Andrès, "Classification models for heart disease prediction using feature selection and pca," *Informatics in Medicine Unlocked*, vol. 19, p. 100330, 2020.

[148] P. Punithavathi, S. Geetha, M. Karuppiah, S. H. Islam, M. M. Hassan, and K.-K. R. Choo, "A lightweight machine learning-based authentication framework for smart iot devices," *Information Sciences*, vol. 484, pp. 255–268, 2019.

[149] H. Ke, D. Chen, T. Shah, X. Liu, X. Zhang, L. Zhang, and X. Li, "Cloud-aided online eeg classification system for brain healthcare: A case study of depression evaluation with a lightweight cnn," *Software: Practice and Experience*, vol. 50, no. 5, pp. 596–610, 2020.

[150] D. Kaur, M. Sobiesk, S. Patil, J. Liu, P. Bhagat, A. Gupta, and Markuzon N., "Application of Bayesian networks to generate synthetic health data," *Journal of the American Medical Informatics Association*, vol. 28, no. 4, pp. 801–811, 2021.

[151] D. Kaur, R. J. Panos, O. Badawi, S. S. Bapat, L. Wang, and A. Gupta, "Evaluation of clinician interaction with alerts to enhance performance of the tele-critical care medical environment," *International Journal of Medical Informatics*, vol. 139, pp. 104165, 2020.